\documentclass[a4paper,british,a4paper,fleqn,usenatbib,unicode=true]{mnras}
\usepackage[T1]{fontenc}
\usepackage[latin9]{inputenc}
\usepackage{babel}
\usepackage{array}
\usepackage{float}
\usepackage{textcomp}
\usepackage{amsmath}
\usepackage{graphicx}
\usepackage{wasysym}
\usepackage{esint}
\usepackage[authoryear]{natbib}
\usepackage[unicode=true]
 {hyperref}

\makeatletter

\pdfpageheight\paperheight
\pdfpagewidth\paperwidth

\providecommand{\tabularnewline}{\\}
\floatstyle{ruled}
\newfloat{algorithm}{tbp}{loa}
\providecommand{\algorithmname}{Algorithm}
\floatname{algorithm}{\protect\algorithmname}

\newcommand{\NH}{ {N_{\rm{H}}} }
\newcommand{\LX}{ {L_{\rm{X}}} }
\newcommand{\AP}{ {p} }
\newcommand{\PuffedTorus}{{\it{radiation-lifted torus}} }
\makeatother

\usepackage{listings}

\begin{document}
\title[Galaxy-scale obscuration in Active Galactic Nuclei]{Galaxy
gas as obscurer: II. Separating the galaxy-scale and nuclear obscurers
of Active Galactic Nuclei }
\author[Buchner \& Bauer]{
Johannes Buchner$^{1,2}$\thanks{E-mail: johannes.buchner.acad@gmx.com}, 
Franz E. Bauer$^{1,2,3}$
\\
$^{1}$Millenium Institute of Astrophysics, Vicu\~{n}a. MacKenna 4860, 7820436 Macul, Santiago, Chile
\\
$^{2}$Pontificia Universidad Católica de Chile, Instituto de Astrofísica, Casilla 306, Santiago 22, Chile
\\
$^{3}$Space Science Institute, 4750 Walnut Street, Suite 205, Boulder, Colorado 80301
}{

\date{Accepted XXX. Received YYY; in original form ZZZ}
\pubyear{2015}
\label{firstpage}
\pagerange{\pageref{firstpage}--\pageref{lastpage}}

\maketitle
\begin{abstract}The ``torus'' obscurer of Active Galactic Nuclei
(AGN) is poorly understood in terms of its density, substructure and
physical mechanisms. Large X-ray surveys provide model boundary constraints,
for both Compton-thin and Compton-thick levels of obscuration, as
obscured fractions are mean covering factors $f_{\text{cov}}$. However,
a major remaining uncertainty is host galaxy obscuration. In Paper
I we discovered a relation of $\NH\propto M_{\star}^{1/3}$ for the
obscuration of galaxy-scale gas. Here we apply this observational
relation to the AGN population, and find that galaxy-scale gas is
responsible for a luminosity-independent fraction of Compton-thin
AGN, but does not produce Compton-thick columns. With the host galaxy
obscuration understood, we present a model of the remaining, nuclear
obscurer which is consistent with a range of observations. Our \PuffedTorus
model consists of a Compton-thick component ($f_{\text{cov}}\sim35\%$)
and a Compton-thin component ($f_{\text{cov}}\sim40\%$), which depends
on both black hole mass and luminosity. This provides a useful summary
of observational constraints for torus modellers who attempt to reproduce
this behaviour. It can also be employed as a sub-grid recipe in cosmological
simulations which do not resolve the torus. We also investigate host-galaxy
X-ray obscuration inside cosmological, hydro-dynamic simulations (EAGLE,
Illustris). The obscuration from ray-traced galaxy gas can agree with
observations, but is highly sensitive to the chosen feedback assumptions.\end{abstract}


\section{Introduction}

The vast majority of Active Galactic Nuclei (AGN) are obscured by
thick columns of gas and dust. X-ray surveys over the last decade
indicate that $20-40\%$ are hidden behind Compton-thick column densities
($\NH\apprge10^{24}\text{cm}^{-2}$) and of the remaining population,
$\sim75\%$ are obscured with $\NH=10^{22}-10^{24}\text{cm}^{-2}$
\citep[e.g.][]{Treister2004,Brightman2014,Ueda2014,Buchner2015,Aird2015}
at the peak of AGN activity at redshift $z=0.5-3$. An open question
is whether the same gas reservoir is responsible for fuelling the
AGN by accretion onto Supermassive Black Holes (SMBHs), and whether
it itself is affected by AGN activity. To address this, the first
step is to identify the scale at which the obscuring gas resides.
Traditionally, AGN obscuration is associated with the ``torus'',
a nuclear ($\sim10\text{pc}$) structure surrounding the accretion
disk. Many basic questions about this gas reservoir remain to be answered,
including its density, substructure and stability mechanism \citep{Elitzur2006a,Hoenig2013}.
Assuming sampling from random viewing angles, the high fraction of
obscured AGN implies high covering fractions. Turbulent structures
such as winds from accretion disks have been invoked to explain this
\citep{Krolik1988}. However, for the covering fractions to be useful
constraints for torus models, the importance of galaxy-scale gas to
the obscuration has to be estimated. Separating the covering and column
densities from nuclear and galaxy-scale obscurers is the goal of this
work.

Local galaxies exemplify that several scales can contribute to the
obscured columns. The Milky Way gas distribution shows column densities
of $\NH>10^{22}\text{cm}^{-2}$, but only at very low Galactic latitudes
($|b|\apprle2\text{\textdegree}$, \citealp{Dickey1990,Kalberla2005GalNHdist}).
Towards the Galactic Centre, columns with $\NH>10^{24}\text{cm}^{-2}$
can be found in the Central Molecular Zone \citep{Morris1996,Molinari2011}
and in the equivalent central zones of nearby AGN host galaxies \citep{Prieto2014}.
Also the obscuration in the AGN host galaxy NGC~1068 is clearly nuclear
(in this work: $\sim100\text{pc}$ or smaller), because its Compton-thick
column is observed in a face-on galaxy \citep{Matt1997a}. On the
other hand, many nearby, obscured AGN are hosted in edge-on galaxies
\citep{Maiolino1995}, which suggests that dust-lanes may also be
important obscurers \citep[see also][for galactic optical/infrared extinction]{Goulding2009}.
Hence \citet{Matt2000} proposed a two-phase model for the obscuration
of AGN: a central, nuclear obscurer which provides Compton-thick obscuration,
and the host galaxy, which provides mildly obscured lines of sight. 

However, it has remained difficult to quantifying the covering fractions
of host galaxy-scale gas. The Milky Way and local galaxies are limited
in their use as templates for the high-redshift universe, specifically
at peak SMBH growth ($z=0.5-3$; e.g., \citealp{Ueda2003,Aird2010}).
At that time, the gas content in galaxies was probably higher, as
indicated by molecular gas measurements \citep[e.g.][]{Tacconi2013}.
To observationally decouple the galaxy-scale and nuclear X-ray obscurer,
we need to go beyond a single, central source. 

This work builds on observational results from Paper~I of the obscuring
column distribution of galaxy-scale gas alone. This stellar-mass dependent
result was derived from the X-ray spectra of an unbiased Gamma Ray
Burst (GRB) sample. In Section~\ref{sec:Methodology} we present
our computation of transferring these results from the GRB host population
to the AGN host population, taking into account the different stellar
mass distribution. Section~\ref{sec:Results} presents our prediction
of the galaxy-scale obscurer alone and compared with obscured fractions
from AGN surveys. Independently, Section~\ref{sec:Cosmological-simulations}
looks into simulated galaxies in hydro-dynamic cosmological simulations.
These also provide predictions for the amount of gas inside galaxies,
from which we derive obscured fractions using ray-tracing. We discuss
the implications and limitations of our two approaches in Section~\ref{sub:Galaxy-scale-obscuration-for}.
We then subtract the galaxy-scale obscurer from the observed obscuration
and reveal the remaining, nuclear obscurer in Section~\ref{sub:Ldep}.
Its luminosity-dependent behaviour is analysed in detail, for which
we present a model in Section~\ref{sub:PuffedTorus}.

\section{Methodology}

\label{sec:Methodology}Our goal is to predict the fraction of obscured
AGN from the obscuration of host galaxy-scale gas alone, i.e. without
nuclear obscuration (the torus and central molecular zones). In this
fashion we will be able to separate the large-scale and small-scale
obscurer. In Paper I we established a relation between the distribution
of X-ray absorbing column densities, $\NH$, in galaxies and the stellar
mass of the galaxy. It follows approximately a log-normal distribution
around 
\begin{equation}
\NH=10^{21.7}\text{cm}^{-2}\times\left(M_{\star}/10^{9.5}M_{\odot}\right)^{1/3}\label{eq:NHM-rel}
\end{equation}
with standard deviation $\sigma=0.5\,\text{dex}$. This $\NH-M_{\star}$
relation was derived using an unbiased sample of long Gamma Ray Bursts
(LGRBs). Because the obscuration is host mass dependent, the obscurer
is arguably the host galaxy itself. In Paper I we show that modern
cosmological hydro-dynamic simulations reproduce absorption by galaxy-scale
metal gas and predict $\NH-M_{\star}$ relations very similar to Equation~\ref{eq:NHM-rel}. 

We now apply this relation to the AGN host galaxy population to estimate
host galaxy obscuration. This relation was derived from actively star
forming galaxies, therefore a caveat is that results may slightly
over-represent the galaxy gas present in AGN host galaxies which have
average \citep{Rosario2011,Rosario2013,Santini2012} or even below-average
\citep{Mullaney2015} star formation rates. The major benefit of using
this relationship is that it is based on the same observable as AGN
obscured fractions, namely the photo-electric absorption of X-rays.
Paper I also investigated the host galaxy metallicity bias of LGRB
and concluded that it has a negligible effect on the obscurer. Local
galaxies are also shown there to follow the $\NH-M_{\star}$ relation.
In Section~\ref{sec:Cosmological-simulations} we further test whether
our assumptions hold using simulations.

We begin with the stellar mass function (SMF) of the galaxy population.
Its shape $P(M_{\star}|z)$ is approximately a Schechter function
and was measured by e.g. \citet{Muzzin2013} and \citet{Ilbert2013}
out to $z\sim4$. Then we populate the galaxies with AGN. The occupation
probability $P(\text{AGN}|M_{\star},z)$ has been measured by \citet{Aird2012}
and \citet{Bongiorno2012} primarily for the redshift interval $z=0.5-2$.
More accurately, these authors measure the specific accretion rate
distribution (SARD), $P(\LX|M_{\star},z)$, where $\LX$ is the $2-10\,\text{keV}$
X-ray luminosity. They find factorised powerlaw relationships of the
form $P(\LX|M_{\star},z)=A\cdot L_{{\rm X}}^{\gamma_{L}}\cdot M_{\star}^{\gamma_{M}}\cdot(1+z)^{\gamma_{z}}$.
At the highest luminosities, an Eddington limit is required to explain
the steep decrease at the bright end of the luminosity function \citep{Aird2013}.
In this work, however, we focus on the $\LX=10^{42-45}\text{erg/s}$
luminosity range and thus use only the observed relation. The final
ingredient is the obscuring column density distribution $P(\NH|M_{\star})$,
which is given by Equation \ref{eq:NHM-rel} as a log-normal distribution.
We assume that the galaxy-scale gas is independent of nuclear activity
for individual galaxies.

The obscured fraction can then be simply computed by Monte Carlo simulations.
Analytically we can put the three distributions together as

\begin{equation}
P(\NH,\LX,M_{\star}|z)=\underbrace{P(M_{\star}|z)}_{\text{SMF}}\times\underbrace{P(\LX|M_{\star},z)}_{\text{SARD}}\times\underbrace{P(\NH|M_{\star})}_{\text{Eq 1}}\label{eq:combination}
\end{equation}
After inserting the factorised powerlaw relationship of the SARD,
the result has the form

\begin{equation}
P(\NH,\LX,M_{\star}|z)=A\times L_{X}^{\gamma_{L}}\times f(M_{\star}|z)\times P(\NH|M_{\star}).
\end{equation}
The SARD implies that while the absolute probability of finding an
AGN is a function of luminosity, the mass distribution is independent
of luminosity. That is, at every luminosity (not considering the Eddington
limit), the same mix of host stellar masses contribute. Therefore,
the implication is that the galaxy-scale gas obscuration is the same
at all AGN luminosities. An Eddington-limit-like effect, i.e. the
suppression of low-mass galaxies at the bright end, in combination
with the $\NH-M_{\star}$ relation should induce a positive correlation
between luminosity and obscured fraction. Instead, a negative correlation
is observed: bright AGN are less frequently obscured \citep{Ueda2003,Hasinger2005,LaFranca2005,Ebrero2009,Ueda2014,Buchner2015,Aird2015}.
We therefore retain the simple, luminosity-independent formalism.

We adopt an AGN definition of $\LX>10^{42}\text{erg/s}$. The frequency
distribution of column densities $\NH$ for the AGN population is
then computed by integrating over stellar mass and luminosity:

\begin{equation}
P(\NH|z)=\int\int_{10^{42}\text{erg/s}}^{\infty}P(\NH,\LX,M_{\star}|z)\,d\LX\,dM_{\star}.
\end{equation}

\begin{figure*}
\begin{centering}
\includegraphics[width=0.95\textwidth]{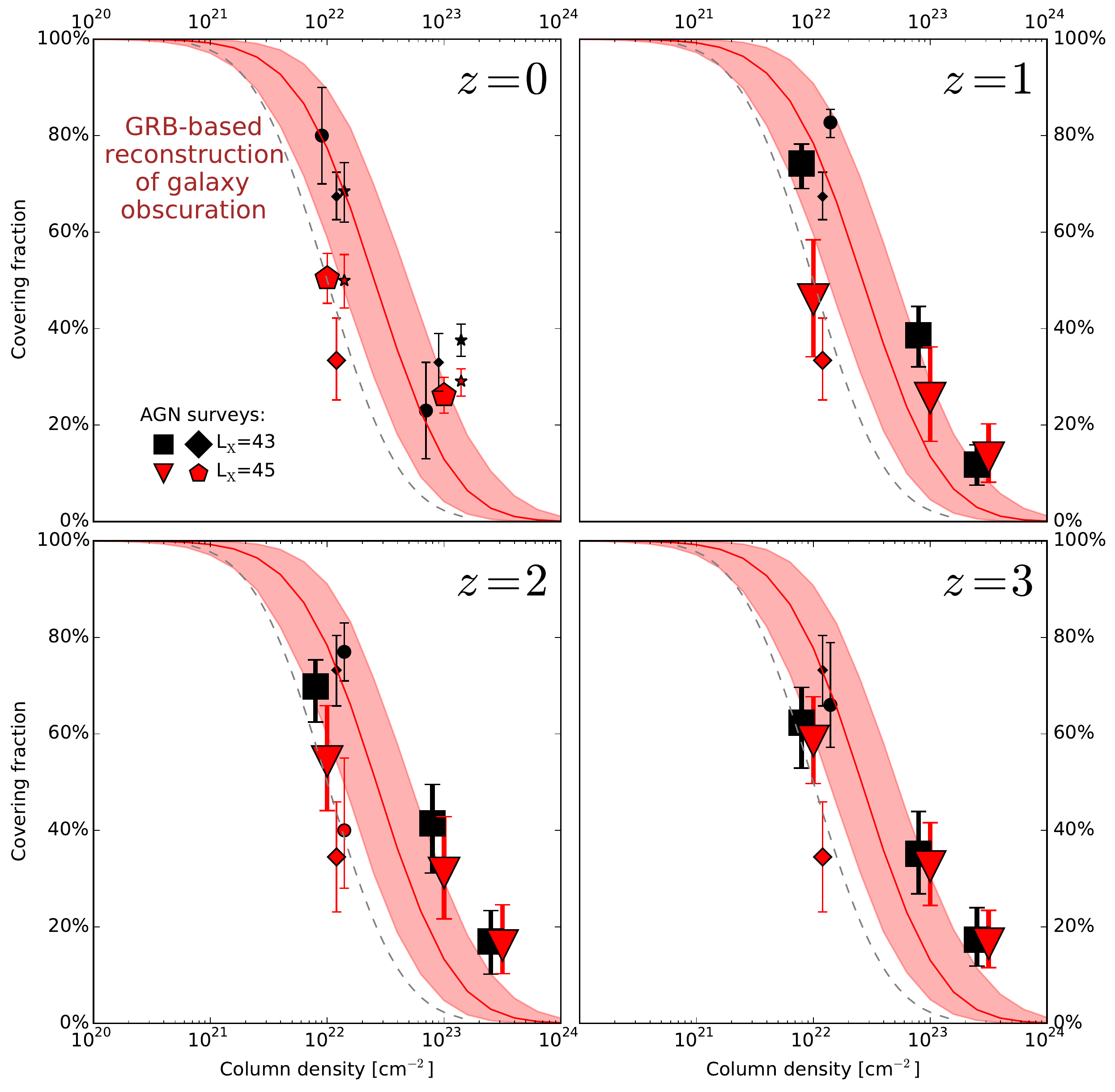}
\par\end{centering}

\caption[Galaxy-scale obscuration of AGN]{\label{fig:gasobscprofilez_obs}Galaxy-scale obscuration of AGN.
At various redshift intervals we show the fraction of AGN (y-axis)
that is covered by a given column density $\NH$ (x-axis). The red
curve indicates our derivations based on GRB tomography with uncertainties
shown in red shading. Galaxy-scale obscuration is negligible for column
densities of $10^{24}{\rm cm}^{-2}$, but is an important contributor
to the AGN fraction at $\NH\approx10^{22-23}\,\text{cm}^{-2}$. Data
points represent intrinsic (flux bias-corrected) obscured fractions
in the Compton-thin population from AGN surveys: Bright AGN ($L(2-10{\rm keV})\approx10^{45}{\rm erg/s}$,
red error bars) show lower obscurations than faint AGN ($L(2-10{\rm keV})\approx10^{43}{\rm erg/s}$,
black error bars). The data points are diamonds, circles, star small
symbols \citep[respectively]{Ueda2014,Aird2015,Ricci2016} and pentagons
and squares/triangles larger symbols \citep[respectively]{Burlon2011,Buchner2015}.
Because AGN also have a nuclear obscurer, these should be regarded
as upper limits for galaxy gas obscuration (red shading). Arguably
below all data constraints is a normal distribution around $10^{22}{\rm \,cm}^{-2}$
(dashed grey curve), which is kept constant across panels.}
\end{figure*}
Finally the obscured AGN fraction is the cumulative distribution,
i.e. the fraction of AGN hidden behind a certain column density threshold
$\NH$ or higher: 
\begin{equation}
f_{\text{cov}}(>\NH)=\int_{\NH}^{\infty}P(\NH'|z)d\NH'.\label{eq:fcov}
\end{equation}
Into the calculation of $f_{\text{cov}}$ we propagate the uncertainties
from the obscuration relation for Paper~I. We consider two SMF (\citealp{Muzzin2013}
and \citealp{Ilbert2013}) and two SARD measurements (\citealp{Aird2012}
and \citealp{Bongiorno2012}), to incorporate systematic uncertainties.
To summarise, we rely only on observational relations to predict the
obscuration of the AGN population by galaxy-scale gas.

\section{Results}

\label{sec:Results}

The obscured fraction of the AGN population from combining the observed
relationships is shown in Figure~\ref{fig:gasobscprofilez_obs}.
Each panel represents a specific redshift. The red curve shows our
fraction of obscured AGN (y-axis) for a given column density $\NH$
(x-axis). Red shading shows the uncertainties ($2\sigma$), which
are dominated by the $\NH-M_{\star}$ relation ($\pm0.2\text{\,dex}$),
while our marginalisation over different SARD and SMF moves the $\NH$
distribution by less than $0.08\,\text{dex}$ and $0.04\,\text{dex}$,
respectively (not depicted).

Firstly, galaxy-scale gas does not provide Compton-thick column densities
($\NH>\sigma_{T}^{-1}=1.5\times10^{24}{\rm cm}^{-2}$). This is because
massive galaxies which reach those densities are rare and therefore
represent a negligible fraction of the AGN population. In contrast,
the observed fraction of Compton-thick AGN is $\sim38\%$ \citep[e.g.][]{Buchner2015,Ricci2016,Aird2015}.
We can conclude that Compton-thick obscuration is always associated
with the nuclear region. An alternative, theoretical argument based
on the total metal gas mass present in galaxies is laid out in Appendix~\ref{sub:semi-analytic}
and arrives at the same conclusion.

We therefore focus on the Compton-thin obscurer. This step implicitly
assumes that the Compton-thick nuclear obscurer is randomly oriented
with respect to the galaxy, in accordance with chaotic accretion \citep{King2006}.
We now compare with measurements of the obscured fraction of Compton-thin
AGN from surveys. When these fractions are treated as covering fractions,
they contain both galaxy-scale obscuration and nuclear obscuration.
Therefore the data points should always be understood as upper limits
to the galaxy-scale gas obscuration. Figure~\ref{fig:gasobscprofilez_obs}
shows results from surveys at the peak of the accretion history ($z=1-3$
panels, \citealt{Ueda2014,Buchner2015,Aird2015}) and surveys which
include the local Universe \citep{Burlon2011,Ueda2014,Aird2015,Ricci2016}.
Results are often given as the intrinsic fraction of Compton-thin
AGN with $\NH>10^{22}\text{cm}^{-2}$, a common definition of ``obscured''
AGN. When comparing our results with these data points in Figure~\ref{fig:gasobscprofilez_obs},
we find that galaxy-scale obscuration is likely the dominant contributor
at $\NH=10^{22-23}\text{cm}^{-2}$. Two cases are important: (1) the
fraction of obscured AGN at the bright end ($\LX\geq10^{45}\text{erg/s}$),
where the lowest obscured fractions are observed ($f_{\text{bright}}\sim40\%$,
shown as red data points), and (2) at the faint end ($\LX\sim10^{43}\text{erg/s}$),
where the highest obscured fractions are observed ($f_{\text{faint}}\sim70\%$,
shown in black). Our results for galaxy-scale obscuration are, as
discussed in Section~\ref{sec:Methodology}, luminosity-independent.
Because of the large uncertainties, we cannot distinguish whether
galaxy-scale obscuration decreases toward the bright end, or increases
toward the faint end. In Section~\ref{sec:Discussion} we also examine
systematic uncertainties from the enhanced star formation rates of
LGRB hosts compared to AGN hosts. In any case, however, our prediction
indicates that galaxy-scale obscuration is sufficient to explain the
observed fractions of obscured AGN at $\NH=10^{22-23}\text{cm}^{-2}$.

At higher column densities the observed obscured fractions lie systematically
higher than our results from galaxy-scale gas. Therefore a nuclear
obscurer is required to explain this obscuration excess and becomes
dominant at approximately $\NH\approx10^{23.5}\text{cm}^{-2}$. The
shape of the distribution is driven by the Gaussian distribution of
the $\NH-M_{\star}$ relation, the distribution adopted in Paper I
to fit the dispersion of LGRB LOS column densities. Adopting a different
distribution would result in the same width, but may allow flatter
tails and permit more unobscured LOS.

We emphasise that our results are meaningful for the AGN population
-- the obscuration of individual host galaxies is stellar-mass dependent
with substantial variations between individual galaxies (see Equation
\ref{eq:NHM-rel}).

\section{Cosmological hydro-dynamic simulations}

\label{sec:Cosmological-simulations}In this section we assess the
gas content in simulated galaxies. This allows us to validate our
approach and to compare with the predictions for galaxy-scale obscuration
by those simulations. Modern cosmological hydro-dynamic simulations
self-consistently evolve galaxies and their processes (star formation,
gas accretion, supernova and AGN feedback, etc.) in the context of
well-constrained $\Lambda$CDM cosmologies. These simulations use
the initial conditions of the baryon density available in the early
Universe and are tuned to reproduce the local galaxy population. We
consider two state-of-the-art cosmological hydro-dynamic simulations
which also produce realistic galaxy morphologies. These simulations
allow us to look at the spatial distribution of gas inside galaxies.

\subsection{Simulation sets: EAGLE \& Illustris}

The Evolution and Assembly of Galaxies and their Environment (\textbf{EAGLE})
simulation \citep{Schaye2015,Crain2015} reproduces many observed
quantities; it reproduces very well the stellar mass function \citep{Furlong2015a}
and size distribution \citep{Furlong2015} of galaxies as a function
of cosmic time, being calibrated to reproduce these at $z=0$. Further
relevant for this work, it also produces galaxies with realistic galaxy
morphologies \citep{Schaye2015} and gas contents consistent with
observations of CO and HI \citep{Bahe2015} as well as H$_{2}$ \citep{Lagos2015}.
This encourages us to look inside simulated galaxies and assess the
obscuration provided by them. EAGLE includes black hole particles,
which are seeded into dark matter halos exceeding masses of $10^{10}M_{\odot}$.
These black holes are kept near the galaxy centre and may accrete
when gas is nearby, in turn activating a heating feedback mechanism
\citep{Schaye2015}. The strength of EAGLE lies in its minimalistic
subgrid recipes and the systematic exploration of alterations. Besides
the reference model (\texttt{L0100N1504\_REFERENCE}) which we use
primarily, a series of simulations with parameter variations have
been run and analysed. These explore the impact of the type and strength
of supernova feedback, AGN heating and criteria for when stars are
formed \citep{Crain2015}. 

We also consider \textbf{Illustris} \citep{Vogelsberger2014,Vogelsberger2014a},
another hydro-dynamic cosmological simulation. This simulation also
reproduces many observed quantities; most relevant for this work is
that it reproduces the morphology of galaxies, the gas content from
CO observations \citep{Vogelsberger2014,Genel2014}. The Illustris
sub-grid models were chosen in consideration of the stellar mass function,
star formation history and mass-metallicity relation. However, the
weak tuning provides only mediocre agreement with regards to the galaxy
stellar mass function \citep{Schaye2015} and size distribution \Citep{Furlong2015}.
On the positive side, the Illustris simulation is based on the AREPO
hydrodynamics code which has been shown to reproduce galaxy features
well \citep[e.g.][]{Vogelsberger2012,Nelson2013}. The gas particle
resolution in \emph{Illustris} is adaptive, with some cells being
as small as $48\,{\rm pc}$ in the highest resolution simulation (\emph{Illustris-1})
used here, indicating that modern cosmological simulations indeed
resolve galaxies into small sub-structures. Similar to EAGLE, Illustris
includes black holes which are created and kept in the gravitational
potential minimum of galaxies inside halos of mass $M_{\text{DM}}>7.1\times10^{10}M_{\odot}$
\citep{Sijacki2015}. 
\begin{figure*}
\begin{centering}
\includegraphics[width=0.95\textwidth]{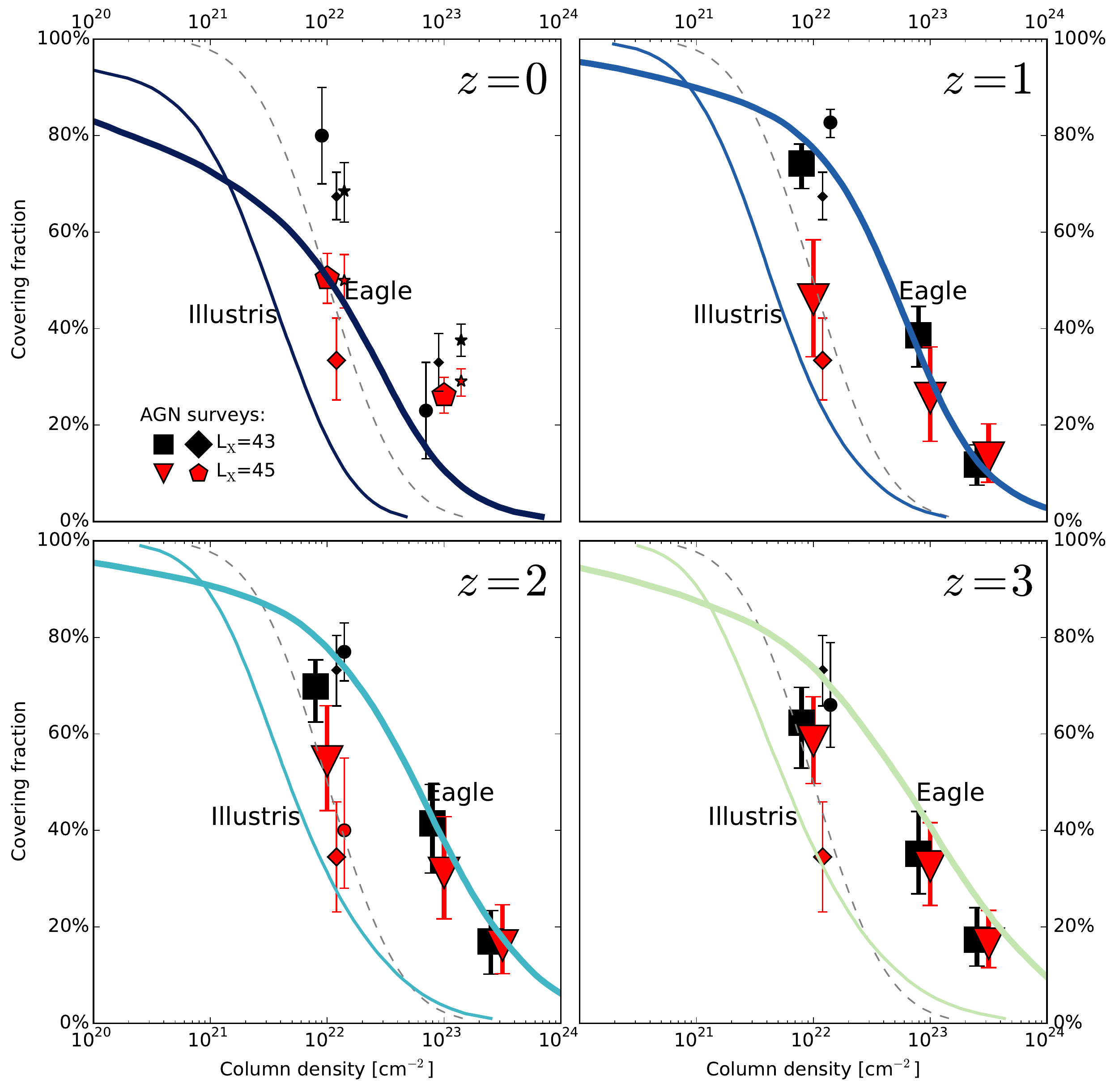}
\par\end{centering}

\caption[Galaxy-scale obscuration of AGN]{\label{fig:gasobscprofilez_sim}Hydro-dynamic cosmological simulation
results for the galaxy-scale gas obscuration of AGN. At various redshift
intervals (panels) we show the fraction of AGN (y-axis) that is covered
by a given column density $\NH$ (x-axis). Curves indicate results
from ray-tracing the metal gas. The EAGLE reference simulation (thick
line) produces thicker column densities than Illustris (thin line).
Data points are the same as in Figure~\ref{fig:gasobscprofilez_obs}.
These show fractions from surveys of bright/faint AGN in triangle/square
symbols respectively. Because AGN also have a nuclear obscurer, these
should be interpreted as upper limits for galaxy gas obscuration.
The dashed grey curve (same as in Figure~\ref{fig:gasobscprofilez_obs})
is kept constant across panels for reference.}
\end{figure*}

\subsection{Ray tracing of gas}

We first investigate the gas distribution in the reference simulations.
We focus on the metal component of gas as O and Fe are, for the relevant
obscuring columns and redshifts discussed here, the most important
elements for photo-electric absorption of X-rays. In galaxy evolution
models, the massive end of the existing stellar population expels
metals into the galaxy. The metal gas produced per stellar mass is
determined by the chosen IMF and the metal yield, with the latter
tuned to reproduce the stellar mass function \citep[e.g.][]{Lu2015}.
The total metal gas mass residing in galaxies further depends on the
chosen feedback models which can expel gas out of the galaxy. The
strength of feedback is also constrained by matching the stellar mass
function and galaxy properties \citep[e.g. the color distribution,][]{Croton2006}.
Typically the metal gas mass inside galaxies follows a $M_{Z}:M_{\star}$
ratio between $1:30$ and $1:100$ in semi-analytic models at $z=0-3$
\citep[e.g.][]{Croton2006,Croton2016}; Plots of these models can
be found in Appendix~\ref{sub:semi-analytic}. The crucial remaining
uncertainty is the arrangement of that gas inside galaxies, as the
concentration of gas defines its column density.

\begin{figure*}
\begin{centering}
\includegraphics[width=0.99\textwidth]{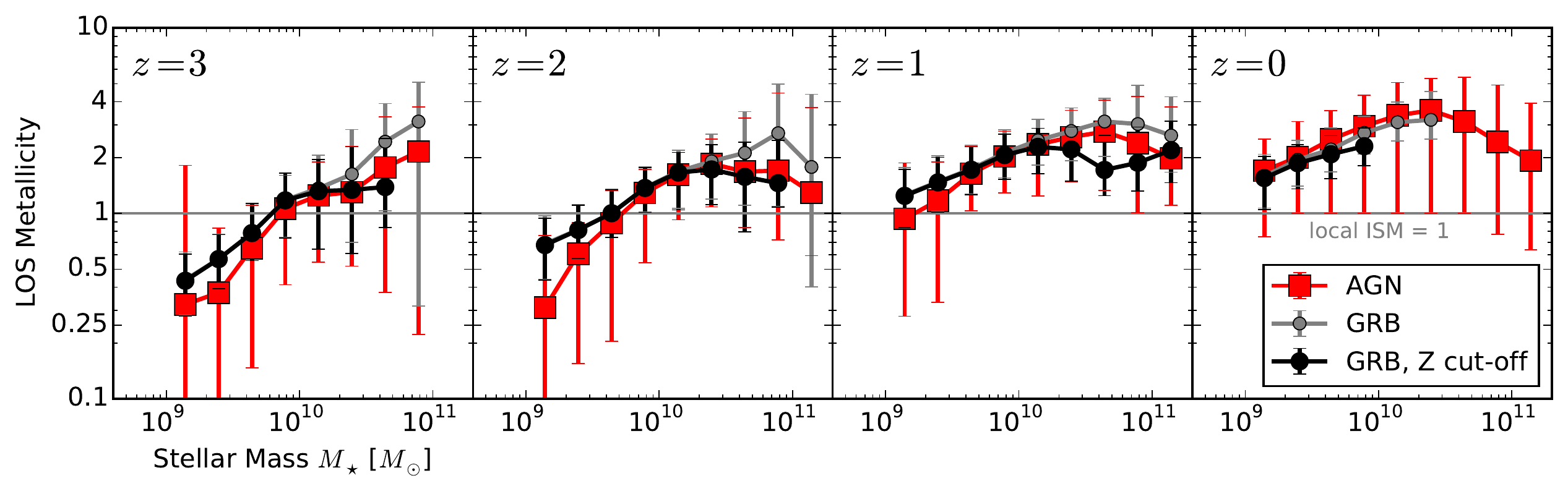}
\par\end{centering}

\caption[LOS column abundances]{\label{fig:Z-sim}LOS column metal abundances. In the EAGLE reference
simulation, hydrogen and metal gas densities were integrated along
random sightlines from the central black hole to the edge of the host
galaxy (AGN, red squares). The ratio is presented relative to local
ISM abundances. For GRBs (grey circles), ray tracing starts from the
hydrogen-densest regions. Black circles are the same, but only allow
galaxies with sub-solar stellar metallicity as GRB hosts. Such a sub-selection
is necessary to reproduce the mass and luminosity distribution of
the observed GRB population \citep{Hjorth2012,Vergani2015,Kruehler2015,Perley2015b}.
Curves present the median, while the errorbars represent the $1\sigma$
scatter in the population. LOS metallicities show a stellar-mass dependence
and evolution over cosmic time, but are very similar for GRBs and
AGN.}
\end{figure*}
We apply ray tracing starting from the most massive black hole particle
of each simulated galaxy (subhalo). From that position, we radiate
in random sightlines any metal gas bound to the subhalo. Each location
on the ray is assigned the density from the nearest gas particle (Voronoi
tesselation)\footnote{Our Voronoi tessellation ray tracing code can be found at \href{https://github.com/JohannesBuchner/LightRayRider}{https://github.com/JohannesBuchner/LightRayRider};
Catalogues of the obscuration of all considered simulated galaxies
are available from the first author on request. }. Finally the density is integrated to a total metal column density.
We then compute an equivalent hydrogen column density distribution
by adopting \citet{Wilms2000} local inter stellar medium (ISM) abundances.
This mimics how $\NH$ is derived in X-ray observations. We adopt
$h=0.7$ and work in physical units at redshifts $z=0,\,1,\,2$ and
$3$. 

Using the covering fractions for each SMBH we compute the obscured
fraction of the AGN population. Because reproducing the luminosity
function from simulated black hole accretion is on-going research
\citep[see][ for Illustris and EAGLE respectively]{Sijacki2015,Rosas-Guevara2016},
we do not rely on the instantaneous accretion rates provided by the
simulations. The effect of adopting these is discussed in Appendix~\ref{sec:subgrid-discuss}.
Instead, for each galaxy we randomly assign a luminosity according
to the SARD of \citet{Aird2012}\footnote{We use host galaxy stellar masses within twice the half-light radius.}
and use only those with $L(2-10\,{\rm keV})>10^{42}\text{erg/s}$.
The last step is repeated 400 times to increase the sample size. With
the column density distribution for each AGN available, we compute
the obscured fraction as a function of column density $\NH$ of the
simulated population.

\subsection{Covering fraction of simulated galaxies}

\label{sub:Results}We present the fraction of AGN showing column
densities larger than a given $\NH$ value in Figure \ref{fig:gasobscprofilez_sim}.
The plot is made in the same fashion as the previous observational
Figure~\ref{fig:gasobscprofilez_obs}. We find that both the EAGLE
(thick curve) and Illustris (thin curve) simulations produce a negligible
number of Compton-thick AGN through host-galaxy obscuration. This
is consistent with the assumption that Compton-thick obscuration is
associated with a nuclear obscuration in the unresolved vicinity of
the black hole. 

For comparison with observations, obscured fractions from AGN surveys
are used, as before in Section~\ref{sec:Results}. Downwards-pointing
triangles in Figure~\ref{fig:gasobscprofilez_sim} indicate constraints
for the obscured fraction of luminous, Compton-thin AGN. Arbitrary
additional nuclear obscuration may be included in them, therefore
they should be interpreted as upper limits to the galaxy-scale obscuration.
We find that the Illustris simulation (thin curve) fulfils these constraints,
as it always produces covering fractions below $40\%$ for columns
of $\NH\geq10^{22}\text{cm}^{-2}$. In contrast, the EAGLE reference
simulation (thick curve) produces an excess of obscured AGN at $\NH=10^{22}\text{cm}^{-2}$
at redshifts $z=1-3$ in Figure~\ref{fig:gasobscprofilez_sim}. This
is in violation of observations even when the higher data point from
low-luminosity AGN is considered. At higher column densities, the
large-scale galaxy gas of the EAGLE simulation produces covering fractions
consistent with the observations, with no need for a nuclear obscurer
up to $\NH=10^{23.5}\text{cm}^{-2}$. In contrast, Illustris galaxies
do not provide column densities of $\NH>10^{23}\text{cm}^{-2}$ and
thus require a nuclear obscurer to explain those observations. At
redshift $z=0$, both EAGLE and Illustris are consistent with the
data points.

To summarise, the covering fractions show a similar behaviour as our
observational results presented in Section~\ref{sec:Results}. Galaxy-scale
obscuration is an important obscurer at column densities $\NH=10^{22}\text{cm}^{-2}$.
The differences between EAGLE and Illustris simulations can be explained
by the sub-grid physics adopted. Appendix~\ref{sec:subgrid-discuss}
investigates sub-grid physics variations in detail. In general, weaker
feedback implementations lead to very high covering factors. Therefore,
Appendix~\ref{sec:subgrid-discuss} concludes that obscured fractions
provide upper limits which may be used to rule out specific feedback
implementations.

\subsection{Metallicity of sightlines to AGN and GRBs}

\label{sub:Metallicity-sim}The metallicity inside galaxies may be
different in GRB star-formation sites and the nucleus of galaxies.
Therefore, we now investigate the hydrogen column density of gas in
the EAGLE\footnote{EAGLE subdivides gas in hydrogen, helium and individual metals, while
Illustris only separates out neutral hydrogen.} simulation and compute the metallicity of sightlines. We note that
\citet{Bahe2015} found good agreement between the EAGLE galaxies
and observations in terms of hydrogen masses and surface densities.
We discuss in Paper I that typical GRB sightlines are observationally
consistent with the metal abundance of the local ISM.

Figure~\ref{fig:Z-sim} shows that the LOS metal abundances are very
similar for AGN and LGRBs. The metal abundance $N_{\text{Z}}/\NH$,
normalised to \citealp{Wilms2000} abundances is presented. For AGN
(red squares), sightlines end at the central black hole. The conditions
and sites of creation for LGRBs are not completely understood, however
they trace the blue light in galaxies \citep{Bloom2002,Fruchter2006}.
Under the simple assumption that LGRBs are created in star-forming
regions of high hydrogen content, we start our ray-tracing from the
region which is densest in hydrogen-gas in each galaxy. The corresponding
sight-line abundances are shown as grey circles in Figure~\ref{fig:Z-sim}.
If we further consider that LGRBs only occur in galaxies with stellar
metallicities below solar, we arrive at the results presented by black
circles. Such a metallicity cut is necessary to reproduce the mass
and luminosity distribution of the observed GRB population, as well
as observed metal emission lines \citep{Hjorth2012,Vergani2015,Kruehler2015,Perley2015b}.
Remarkably, the metallicity cut has little effect on the absorbing
LOS metallicity. At all redshifts and stellar masses, we find similar
metal abundances between AGN and LGRB sightlines. In general, sightlines
average across large parts of the galaxy, and thus are not affected
by local metal variations. This confirms the assumptions of Section~\ref{sec:Results},
where we applied host galaxy obscuration measured in LGRBs sightlines
to the AGN population.

The abundance of metals along sightlines increases with stellar mass
in Figure~\ref{fig:Z-sim}. In massive galaxies, a larger number
of massive stars are available to pollute the ISM and the deeper galaxy
potential can retain more metal gas. Metal abundances also increase
over cosmic time, as metals build up inside galaxies. For AGN, the
SARD skews the stellar mass distribution to the massive end. There,
LOS abundances of about twice the local ISM metallicity are predicted
at all considered redshifts. In contrast, LGRB host galaxies are predominantly
found to have $M_{\star}\approx10^{9}M_{\odot}$ at $z\sim2$ \citep[see e.g.,][]{Schulze2015,Perley2015a,Perley2015b}.
Therefore sub-solar metallicities are to be expected in LGRB afterglow
spectroscopy. For research on the optical absorption of LGRBs by HI,
we refer interested readers to the simulations of \citet{Pontzen2010}.

\section{Discussion}

\label{sec:Discussion}

\subsection{AGN are obscured by their host galaxy}

\label{sub:Galaxy-scale-obscuration-for}Our main result is that the
host galaxy gas provides a luminosity-independent obscurer, for which
we compute covering fractions. This obscurer does not have Compton-thick
column densities, but high covering fractions ($40-90\%$) at $\NH=10^{22}\,{\rm cm}^{-2}$.
These high covering fractions suggest that a substantial part of the
type1/type2 dichotomy is caused by galaxy-scale gas. This is consistent
with the finding of \citet{Maiolino1995} that nearby type2 AGN are
often edge-on galaxies. 

We note that our results tend to produce high covering fractions,
which are only consistent with the measurements because of large uncertainties.
Despite our efforts to mitigate stellar mass and metallicity biases,
we suppose that the use of LGRB host galaxies in Paper~I, may affect
our results when applied to AGN host galaxies: LGRB host galaxies
are actively star forming \citep{Levesque2014,Kruehler2015,Perley2016}
and thus may have slightly more gas than the average AGN galaxy. A
correction in the column density by a factor of 2 would agree well
with data points at all redshifts. This is demonstrated by the dashed
line in Figure~\ref{fig:gasobscprofilez_obs}, which shows a normal
distribution of column densities around $\log\NH/{\rm cm}^{-2}=10^{22}$
with standard deviation $\text{0.5}$. With this potential systematic
over-estimation in mind, a nuclear obscurer is then clearly needed
at $\NH\approx10^{23.5}\,{\rm cm}^{-2}$. In any case, our LGRB-based
results are consistent with the current AGN surveys. 

We evaluate the validity of using LGRBs to probe the galaxy obscuration
of AGN below. Firstly, the stellar mass distributions of LGRBs and
AGN are different. We incorporated this in Section~\ref{sec:Methodology}
by deriving a stellar-mass dependent obscuration of LGRBs, and applied
it to the measured stellar mass distribution of AGN. Secondly, LGRBs
prefer galaxies with sub-solar stellar metallicities, so one may expect
the gas inside them to be of lower metal content than AGN. Furthermore,
the location of AGN in the metal-rich centre of galaxies may also
increase their metal content, in contrast to LGRBs which explode in
star-forming regions. When investigating these effects observationally
in Paper~I and here with simulations in Figure~\ref{fig:Z-sim},
we find that LOS metallicity is very comparable between the AGN and
LGRB population. We speculate that this is because the LOS metallicity
is averaged over long distances of the host galaxies. We emphasise
that the simulations used here reproduce the observed hydrogen and
CO masses in galaxies and simultaneously the galaxy stellar mass function.
Thirdly, most SMBH accretion is thought to occur in galaxy-galaxy
mergers \citep{Sanders1988,Hopkins2006OriginModel}, although evidence
of an elevated merger fraction in X-ray detected AGN is weak at high-redshift
($z>0.5$) \citep{Grogin2003,Grogin2005,Pierce2007,Gabor2009,Kocevski2011,Kocevski2015}.
It has therefore been suggested that only luminous AGN are preferentially
associated with mergers \citep[see e.g.,][]{Treister2012}. Our LGRB-based
results show that host galaxy obscuration is most important at the
luminous end for Compton-thin obscuration. LGRB host galaxies may
be appropriate in this comparison as they are also frequently merging
or distorted systems \citep{Fruchter2006,Wainwright2007}. Finally,
we investigated the obscuration of AGN by host-galaxy gas in the hydrodynamic
simulations in Figure~\ref{fig:gasobscprofilez_sim}. The results
are completely consistent with our GRB-based results: No Compton-thick
columns are predicted for the AGN population, while substantial covering
is found at $\NH\approx10^{22}\,{\rm cm}^{-2}$. However, the obtained
covering fractions depend on the feedback strength, and are distinctly
different for the EAGLE and Illustris simulations; the latter implements
strong feedback and results in lower covering fractions. Appendix~\ref{sec:subgrid-discuss}
investigates this in more detail, and concludes that the fraction
of obscured AGN can be used to rule out specific feedback implementations.
From the aforementioned investigations, our approach of using LGRB
host-galaxy obscuration to constrain the galaxy-scale obscuration
of AGN appears valid.

Our results give, for the first time, constraints on the galaxy-scale
obscurer alone. We use this in the subsequent sections to disentangle
the AGN nuclear obscurer from the galaxy-scale obscurer (Section \ref{sub:Ldep}),
and to describe their behaviour as a function of accretion luminosity,
black hole mass, host galaxy stellar mass and redshift (Section \ref{sub:PuffedTorus}).
Finally, physical effects leading to this behaviour are discussed
in Section \ref{sub:Physical-effects}.

\subsection{The nuclear obscurer is luminosity and mass-dependent}

\label{sub:Ldep}
\begin{figure*}
\begin{centering}
\includegraphics[width=1\columnwidth]{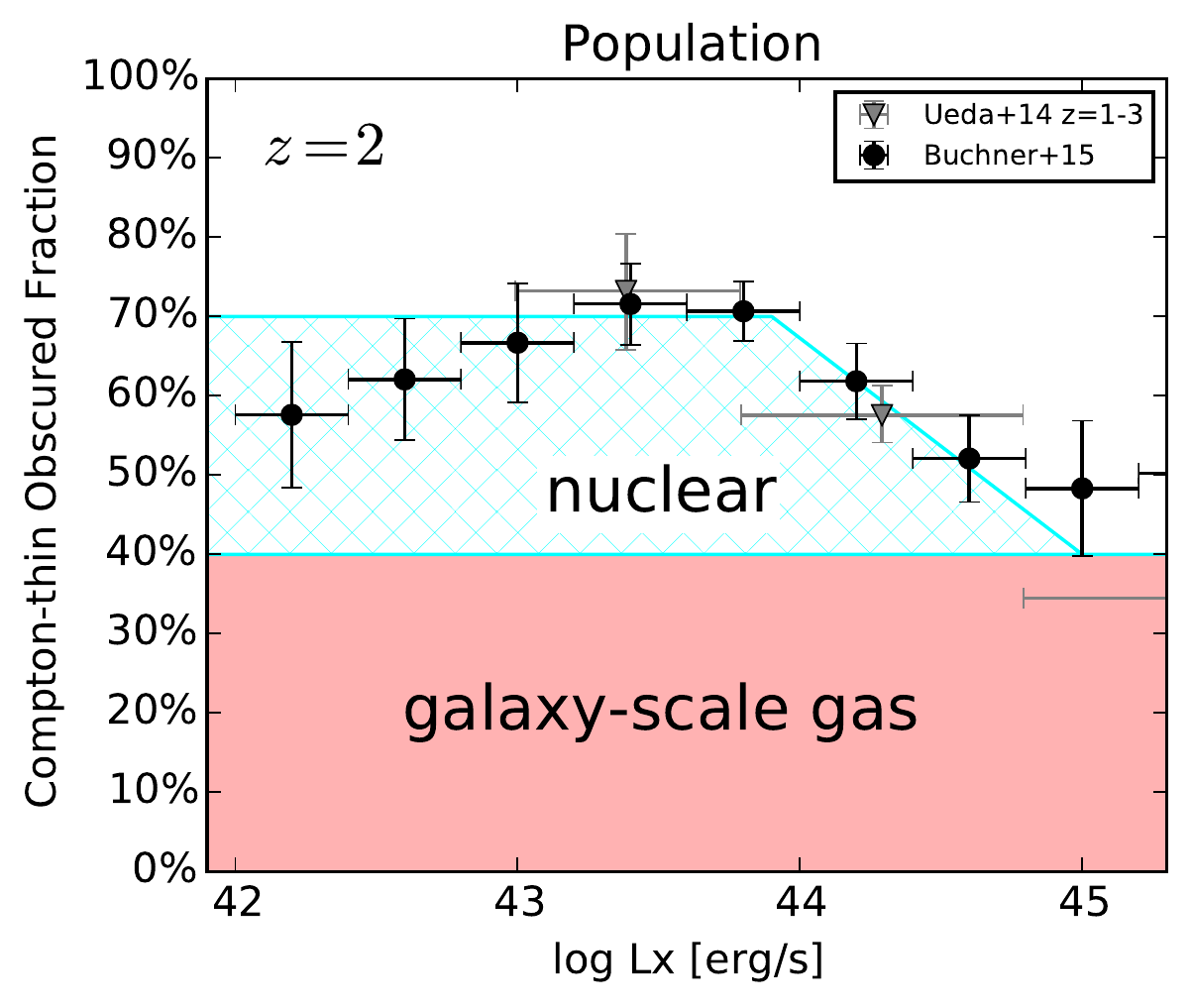}
\includegraphics[width=1\columnwidth]{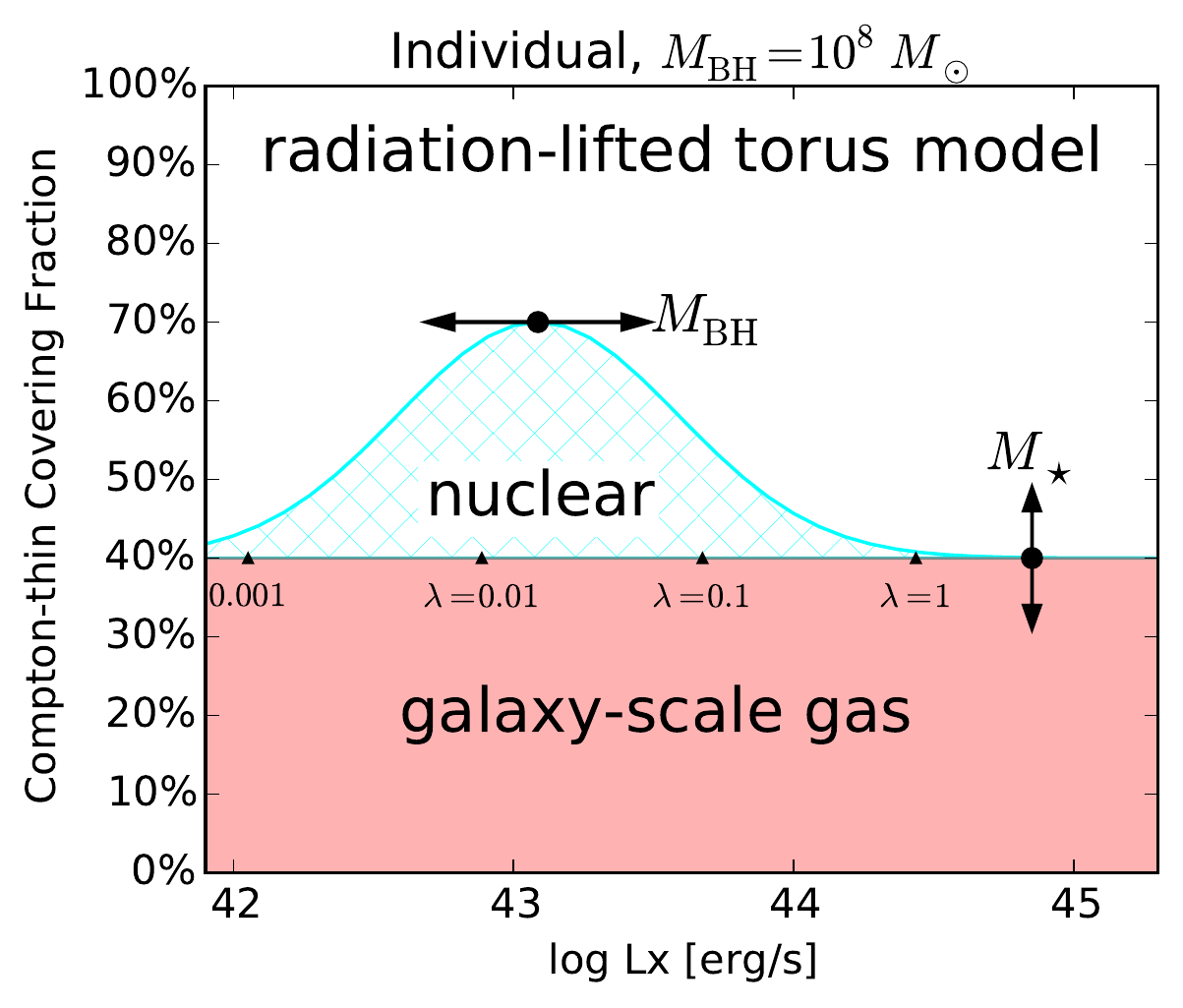}
\par\end{centering}

\caption{\label{fig:L-dep-sketch}\emph{Left panel}: The observed luminosity-dependence
of the obscured fraction of Compton-thin AGN. With the galaxy-scale
gas obscuration constrained and to first order independent of luminosity
(red, see Section~\ref{sec:Results}), we separate out the nuclear,
variable obscurer (cyan, illustrative only) and develop a new model,
the \PuffedTorus model (Section~\ref{sub:PuffedTorus}). \emph{Right
panel}: In the \PuffedTorus model, the nuclear obscuration of individual
sources depends on accretion rate and black hole mass. Here the case
of an individual black hole of mass $10^{8}M_{\odot}$ is shown. The
nuclear obscuration (cyan) is largest at a few percent of the Eddington
rate, and adds to the permanently present, stellar-mass dependent
host galaxy-scale obscuration (red). Code in Listing~\ref{alg:code}.}
\end{figure*}
We can now investigate the behaviour of the nuclear obscurer by subtracting
the galaxy-scale obscuration. In particular, the observed luminosity-dependence
of the obscurer is noteworthy. X-ray surveys consistently find a strong
decline towards high luminosities \citep{Ueda2003,Hasinger2005,LaFranca2005,Ebrero2009,Ueda2014,Buchner2015,Aird2015}
of the obscured fraction $\NH>10^{22}\,{\rm cm}^{-2}$ in the Compton-thin
AGN population. The left panel of Figure~\ref{fig:L-dep-sketch}
shows this decline from $\sim70\%$ to a persistent $\sim40\%$. According
to our results, this can be interpreted as two scales contributing
to the obscuration: a $40\%$ galaxy-scale obscurer, and a luminosity-dependent
nuclear obscurer. The former was constrained in this paper. The latter
is the remainder still needed to fit the observations. According to
this picture, the nuclear obscurer completely disappears toward high
luminosities at $L(2-10\,{\rm keV})\approx10^{44.5}{\rm erg/s}$,
and the only obscurer left is the host galaxy\footnote{aside from the persistent, nuclear, Compton-thick obscurer}.

The nuclear obscurer is however not only luminosity-dependent, but
likely also black hole mass dependent. First evidence of this comes
from noting that the luminosity-dependence is redshift-dependent,
with higher redshifts having higher cut-off luminosities. This has
been found by many works \citep{Ueda2003,Ebrero2009,LaFranca2005,Ueda2014,Aird2015}
by fitting a empirical, parametric model to the relative number density
derived from AGN surveys. \citet{Buchner2015} derived the same result
using a non-parametric method, indicating that this change is indeed
a robust result. In \citet{Buchner2015}, the implications for obscurer
models were discussed in their Section 5.3. As the luminosity dependence
is observed to evolve over cosmic time, a black hole mass dependence
needs to be invoked. They concluded that an Eddington-limited blow-out
of the obscurer could explain the luminosity dependence, because the
Eddington-limit is a function of luminosity and black hole mass. If
the average, currently accreting black hole is more massive at high
redshift than at low redshift, the turn-over luminosity decreases
over cosmic time as observed. Such black hole mass downsizing also
has evidence from optical observations of the black hole mass function
evolution \citep{Schulze2010,Kelly2013a,Schulze2015} and semi-analytic
models which reproduce the evolution of the AGN luminosity function
\citep{Fanidakis2012,Enoki2014,Hirschmann2014}. Recently, \citet{Oh2015}
found evidence that the type-1 fraction in a SDSS selected sample
is both luminosity and black hole mass dependent. Because of the aforementioned
observations, simplistic models which only depend on luminosity such
as the receding torus model \citep{Lawrence1991} have been ruled
out.

We therefore need a luminosity and black hole mass dependent obscurer
model to explain observations. Such an empirical model is presented
in the next section.

\section{Radiation-lifted torus - a nuclear obscurer model}

\label{sub:PuffedTorus}Because neither semi-analytic nor hydro-dynamic
cosmological simulations resolve the nuclear obscurer of AGN, we present
a sub-grid model for post-processing. We construct a model which reproduces
the fraction of obscured AGN as a function of redshift and luminosity
as discussed above. It can also serve as a summary of observational
constraints when exploring physical models for the nuclear obscurer
(see next section).

We assume that a nuclear Compton-thick obscurer covers a fraction
of the SMBH, $f_{\text{CT}}\sim35\%$. This fraction was measured
directly from AGN surveys \citep{Buchner2015}, by matching the soft
and hard X-ray luminosity functions \citep{Aird2015} or by matching
the Compton-thin X-ray luminosity function to the spectrum of the
Cosmic X-ray background \citep{Ueda2014}. Similar fractions have
recently been found in local surveys, e.g. \citet{Ricci2016}. 

\begin{table*}
\caption{\label{tab:Parameters}Parameters of the \PuffedTorus model. The
recommended values are constrained by local observations as described
in Section~\ref{sub:PuffedTorus}. Example code can be found in Listing~\ref{alg:code}.}

\centering{}%
\begin{tabular}{ccc>{\centering}p{7cm}}
Parameter & Symbol & Observed Range & Recommended Value\tabularnewline
\hline 
Compton-thick obscuration & $f_{\text{CT}}$ & 20\% - 40\% & 35\% (fixed)\tabularnewline
Host galaxy obscuration & $f_{\text{gal}}$ & 0\% - 60\% & galaxy-dependent; 40\% population average\tabularnewline
Compton-thin nuclear obscuration & $f_{\text{nuc}}$ & 20\% - 40\% & 30\%\tabularnewline
Luminosity-dependence width & $\sigma$ & 0 - 1 & 0.5\tabularnewline
Reference mass & $M_{L43}$ & $10^{7}$- $10^{8}$$M_{\odot}$ & $10^{7.87}M_{\odot}$\tabularnewline
Mass-dependence & $\AP$ & 0 - 2 & 2/3 (consider also 0, 1, 4/3)\tabularnewline
Intrinsic luminosity ($2-10\,{\rm keV}$) & $L$ & (dynamic) & Determined from accretion rate using bolometric corrections of \citet{Marconi2004}\tabularnewline
\end{tabular}
\end{table*}
We propose that the remaining Compton-thin sky is obscured by two
components: galaxy-scale gas and a nuclear Compton-thin obscurer.
We adopt the following formula for the covering fraction\footnote{exceeding $\NH=10^{22}\text{cm}^{-2}$}:

\begin{eqnarray}
f_{\text{cov}} & = & f_{\text{gal}}(M_{\star})+f_{\text{nuc}}\cdot\exp\left\{ -\frac{1}{2\sigma^{2}}\left(\log\frac{L}{L_{\text{peak}}}\right)^{2}\right\} \label{eq:lumdep}
\end{eqnarray}
The luminosity-independent obscuration, $f_{\text{gal}}$, is in general
a function of host galaxy properties, but on average $\sim40\%$.
In hydro-dynamic simulations the host galaxy obscuration it can be
derived for individual galaxies through ray-tracing (see Section~\ref{sec:Cosmological-simulations}).
Otherwise it can be calculated given a stellar mass using Equation~\ref{eq:NHM-rel}. 

The nuclear obscured fraction is luminosity-dependent. We chose a
Gaussian form which requires fewer parameters than a linear decline
\citep[e.g.][]{Ueda2014}. The obscuration of Equation~\ref{eq:lumdep}
reaches a average maximum of $\sim70\%$ at luminosity $L_{\text{peak}}$
(see Figure~\ref{fig:L-dep-sketch}), and we therefore suggest $f_{\text{nuc}}=30\%$.
In other words, of the sky seen from the black hole $40\%$ is covered
by the host galaxy, and of the remainder the nuclear obscurer covers
half. The obscured fraction declines towards both bright and faint
ends to $f_{\text{gal}}$, with $\sigma$ giving the characteristic
width of the transition. Evidence of the low-luminosity decline has
been found in surveys of the local Universe \citep{Burlon2011,Brightman2011b}
and at high redshifts \citep{Buchner2015}.

The luminosity where the obscured fraction is highest, $L_{\text{peak}}$,
is in turn a function of mass (see Section~\ref{sub:Galaxy-scale-obscuration-for}):

\begin{align}
L_{\text{peak}} & =10^{43}\text{erg/s}\cdot\left(\frac{M_{\text{BH}}}{M_{L43}}\right)^{\AP}
\end{align}
Here, at $M_{L43}$ the distribution peaks at $L(2-10\,{\rm keV})=10^{43}\text{erg/s}$,
which is $L_{\text{bol}}=10^{10.7}L_{\odot}$ using the conversion
of \citep{Marconi2004}. The black hole mass-dependence is defined
by the $\AP$ parameter. At $\AP=0$, the \PuffedTorus is mass-independent,
corresponding to a strict luminosity-dependent unified model. At $\AP=1$,
it is only Eddington-rate dependent. The parameters $\AP$, $M_{L43}$
and $\sigma$ are not known apriori. For mass-scaling, $\AP=2/3$
is suggested from the theoretical works on the obscurer by \citet{Elitzur2016}
and \citet{Wada2015}, however a wide range (e.g. $\AP=0-2$) may
be considered \citep[see also][]{Hoenig2007}. We derive fiducial
values for the other two parameters using the \emph{Swift}-BAT survey
of local AGN. That survey reported a mean mass of $\log M_{\text{BH}}/M_{\odot}=7.87$,
a standard deviation of $0.66\,{\rm dex}$ and a skew towards low
masses \citep{WinterBAT2009}. Adopting an appropriate skewed normal
distribution with skew parameter $-10$ (tail to low masses) around
$M_{L43}=10^{7.87}M_{\odot}$, we find that $\sigma\approx0.5$ approximately
reproduces the width of the obscured fraction function reported in
\citet{Burlon2011}, and peaks at $\LX=10^{43}\text{erg/s}$. 

Table \ref{tab:Parameters} lists the parameters of the \PuffedTorus
model, with recommended typical values. We emphasise that the observations
pertaining to the redshift evolution have not been used when constructing
our model. Cosmological simulations using the \PuffedTorus model
can thus compare be against those \citep[e.g.][]{Buchner2015}. For
simulations, Listing~\ref{alg:code} shows Python code which generates
column densities.

The right panel of Figure~\ref{fig:L-dep-sketch} illustrates the
behaviour of the \PuffedTorus model. The obscured fraction undergoes
a luminosity and mass-dependent peak, where the Compton-thin medium
is extended and occupies a substantial fraction of the sky. Owing
to the host galaxy, a constant fraction is present at all luminosities
and masses. The Compton-thick fraction here has been assumed to be
constant, although we note that \citet{Ricci2016} claims a luminosity-dependence
of the Compton-thick fraction.

\subsection{Predictions of the \PuffedTorus model}

\label{sub:Predictions}
\begin{figure}
\includegraphics[width=0.99\columnwidth]{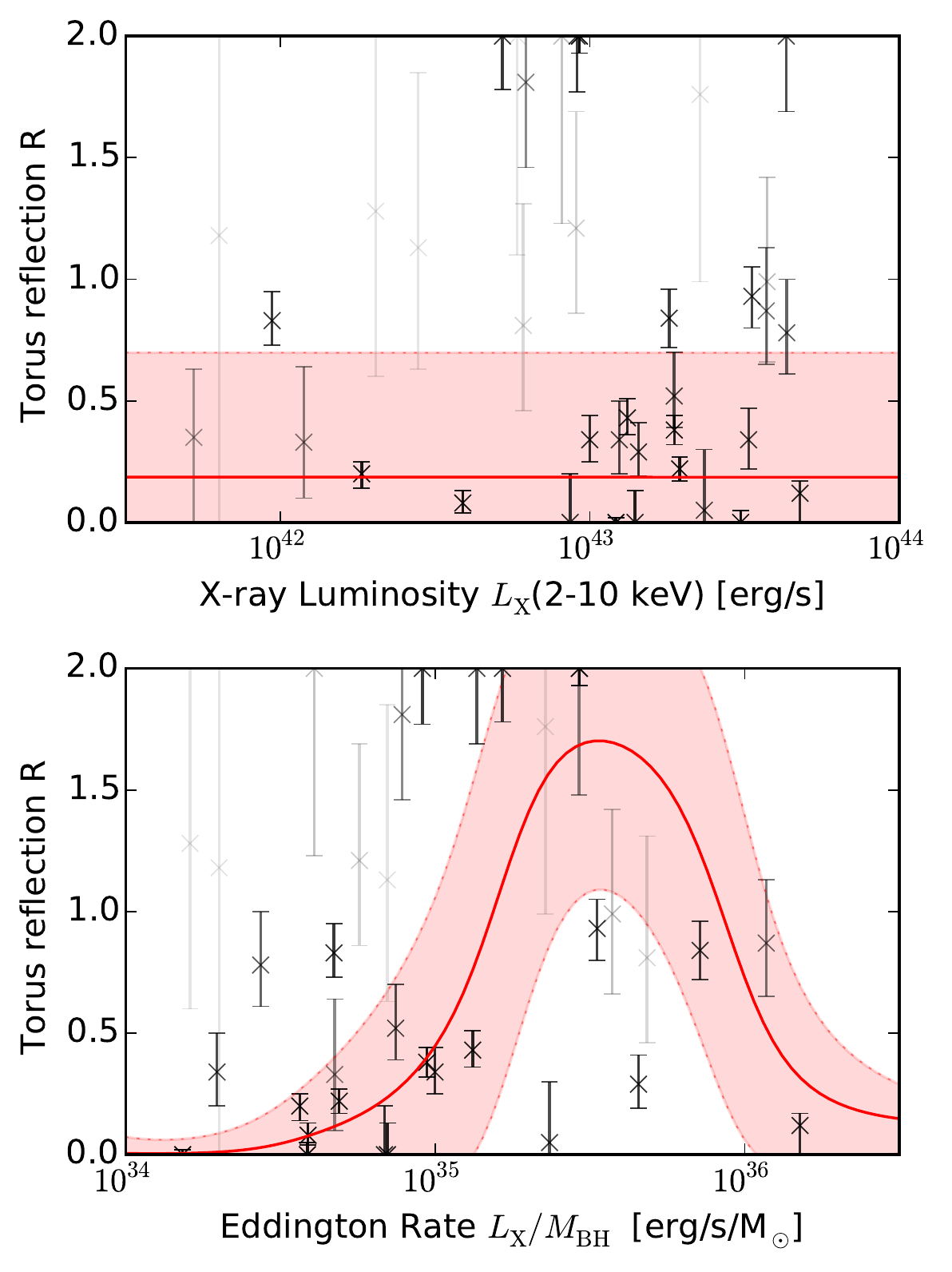}

\caption{\label{fig:torus-reflection-Mdot}The Compton reflection strength,
a proxy for the covering of the torus, as a function of accretion
luminosity (\emph{top panel}) and Eddington rate (\emph{bottom panel}).
Data points are taken from \citet{Kawamuro2016a} and shaded proportional
to the error size to highlight those with tighter constraints. Non-parametric
data smoothing (red curve with $1\sigma$ uncertainties as red shading)
highlights that at intermediate Eddington rates the reflection strength
appears elevated, while there is no clear trend with X-ray luminosity.}
\end{figure}

The \PuffedTorus model predicts that the covering fraction of individual
AGN depends to first order on Eddington rate, not luminosity alone.
\citet{Kawamuro2016a} recently analysed a sample of local ($z<0.1$)
\emph{Swift}-BAT Compton-thin obscured AGN ($\log\NH=10^{22-24}\,{\rm cm}^{-2}$)
with broad-band \emph{Suzaku} observations. They report that the Compton
reflection component in the hard X-ray spectrum varies weakly with
luminosity (their Figure~7, left panel). They use Compton reflection
as a proxy of the covering fraction and argue for a luminosity-dependent
obscurer covering. We plot their data in the top panel of Figure~\ref{fig:torus-reflection-Mdot}.
Overplotted we show a non-parametric fit using a Gaussian kernel density
estimator. Instead of choosing bins manually, this procedure provides
an optimal smoothing length by leaving one data point out in turn,
predicting it from the remaining data, computing the prediction error
and then decreasing the sum of these errors by varying the smoothing
width. In the shown case, a wide kernel is optimal, which indicates
that no strong trend is found, and the data scatters around a mean
value without a strong trend. In other words, luminosity is not effective
in predicting the reflection strength of individual AGN.

If we consider Eddington rate as the driver instead, we find a dependence,
as shown in the lower panel of Figure~\ref{fig:torus-reflection-Mdot}.
While there is substantial scatter in the data, the reflection strengths
of well-constrained objects are high only for intermediate Eddington
rates and low otherwise. The non-parametric fit (red curve) highlights
this. Such a observed bulk increase is rarely expected, as randomly
assigning Eddington rates shows ($p\approx0.05$). Therefore, Figure~\ref{fig:torus-reflection-Mdot}
demonstrates that the evidence for Eddington rate-dependent covering
of individual AGN is stronger than for luminosity-dependent covering.
We caution that the presented evidence is still relatively weak as
the sample of \citet{Kawamuro2016a} is heterogeneously selected and
the Compton reflection strength $R$ is only a crude proxy for the
covering of the obscurer. Upcoming \emph{Swift}-BAT studies with detailed
spectral analysis, completeness corrections and more black hole mass
measurements (C. Ricci et al. in prep) will be in a better position
to confirm or reject the \PuffedTorus model.

The fraction of infrared reprocessed emission has been used to measure
the covering factor of individual AGN. To first order, the obscurer
covering fraction is the ratio of dust infrared re-radiated luminosity
to bolometric illuminating luminosity. Surveys employing this method
\citep[e.g.][]{Maiolino2007,Lusso2013} typically find fractions between
$40\%$ and $75\%$ (luminosity-dependent). These fractions were corrected
for anisotropic illumination and emission \citep[from][see their Figure 13]{Stalevski2016}.
These measurements would include emission from the Compton-thick obscurer
and the nuclear, Compton-thin obscurer. With our assumed fiducial
values we obtain $35\%$ ($f_{\text{CT}}$) to $75\%$ ($f_{\text{CT}}+f_{\text{nuc}}$)
and therefore also agree with those observations. Infrared studies
remain however difficult to use as a constraint, as the entire infrared
SED has to be constrained for each object \citep{Netzer2016} and
covering factors have to be corrected based on uncertain model geometries
\citep{Stalevski2016}.

\subsection{Physical obscurer processes}

\begin{figure*}
\begin{raggedright}
\includegraphics[width=1\textwidth]{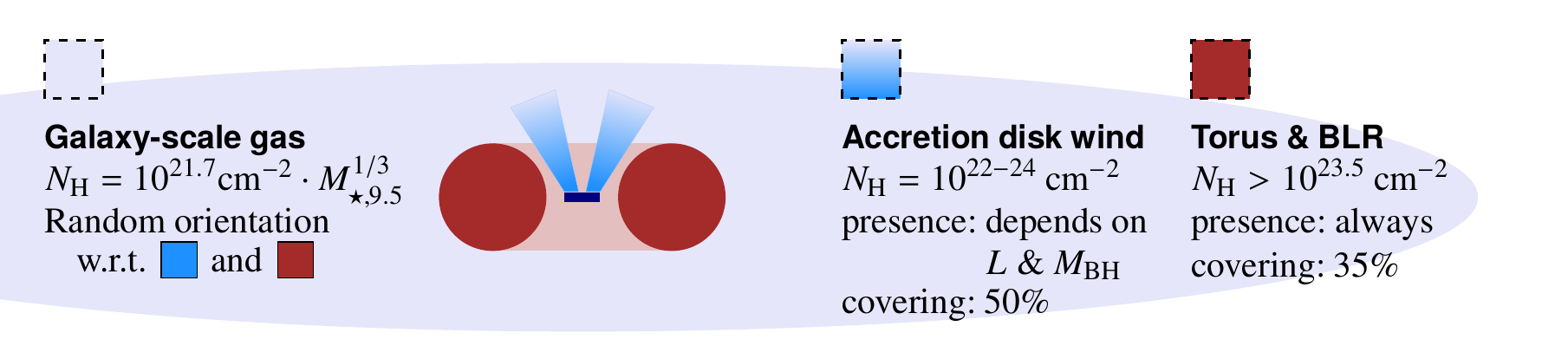}
\par\end{raggedright}

\caption{\label{fig:Cartoon}Cartoon of the three known obscurer components:
Host galaxy obscuration (light blue), luminosity-dependent Compton-thin
nuclear obscuration (blue) and Compton-thick nuclear obscuration (red).
In this illustration, the Compton-thin obscurer is provided by a clumpy
accretion disk wind, while the Compton-thick obscurer is a clumpy
donut-like structure. Both are embedded in the host galaxy with random
orientation. This is one physical scenario consistent with the proposed
\PuffedTorus model, other possibilities are outlined in Section~\ref{sub:Physical-effects}.}
\end{figure*}
\label{sub:Physical-effects}Physical processes giving rise to the
luminosity and mass-dependent behaviour cannot be discussed rigorously
within the scope of this paper. However, we point out a few key results
from recent theoretical works. Otherwise we refer to the review of
\citet{Hoenig2013} which discusses current torus modelling approaches. 

Strong unified obscurer models require that in every object an increase
in the accretion luminosity anti-correlates with the extent of the
obscurer. We take note of the analytic wind model formalism described
in \citet{Elitzur2016} and of the radiation-driven fountain model
by \citet{Wada2015}. Both models produce a vertically extended obscurer
structure as a function of luminosity and mass. At very low luminosities,
radiation is not sufficient to puff up the obscurer. In the hydro-dynamical
simulations of \citet{Wada2012} very high accretion luminosities
are associated with strong outflows, which suppress the vertical extent
of the obscuring structure as they occupy larger angles. The cartoon
of Figure~\ref{fig:Cartoon} illustrates such an accretion disk wind
scenario in combination with host galaxy obscuration and a Compton-thick
torus for the three, distinct obscuring components. The opening angles
shown are approximately correct. For visualisation purposes we have
shown smooth gas distributions, while in reality the obscurer is thought
to be clumpy \citep[see e.g.,][and references therein]{Markowitz2014}.
To date, models of the nuclear obscurer largely lack observational
constraints. Our \PuffedTorus model summarises observational boundary
constraints of covering fractions and column densities for how a physical
``torus'' model should behave.

\citet{Sazonov2015} invoked anisotropic X-ray emission to explain
the luminosity-dependence of the obscurer. If the X-ray emission is
more luminous towards the less-obscured poles while less luminous
towards the more obscured plane, in a coordinate system defined by
the accretion disk. They argue that such anistropy is expected from
accretion disks. Inverse Compton scattering of UV accretion disk photons
in a hot electron plasma near the accretion disk is thought to produce
the X-ray spectrum of AGN \citep{Katz1976,Sunyaev1980,Sunyaev1985}.
Because several such scatterings are needed, it is unclear if the
anistropy would remain \citep[depends on the plasma geometry, see e.g.][]{Yao2005}.
Anisotropic X-ray emission would also predict that AGN with detected
maser disks, which are believed to be viewed edge-on, should have
higher observed Eddington ratios, and thus steeper X-ray spectral
slopes, while this is not observed \citep{Brightman2016}. We also
note that the scenario proposed by \citet{Sazonov2015} naturally
predicts that the luminosity-dependence of the obscuration should
evolve over time, in lock-step with the evolution of the knee of the
luminosity function. Overall, this scenario deserves further study
and can be formally treated as another model of the nuclear obscurer.

A further, commonly overlooked aspect is the triggering mechanism
of AGN. Faint AGN are much more common than luminous AGN \citep[e.g.][]{Barger2005,Ueda2003,Aird2010},
a fact that needs to be explained with the triggering and light-curve
of accretion events. Galaxy-galaxy mergers are today thought to be
the main trigger of luminous AGN activity and efficient SMBH growth.
This is because SMBHs need to accrete substantial fractions of their
host galaxy gas (as shown by scaling relations) within durations comparable
to the dynamical timescales of galaxy centres \citep{Somerville2008}.
In the time-line proposed by Hopkins et al, the luminous phase occurs
in relatively late merger phases. Luminous AGN stop further infall
by radiation pressure and quickly reduce their column densities \citep{Hopkins2006OriginModel}.
In contrast, the faint AGN population is suggested to be dominated
by periods before and after peak accretion \citep{Hopkins2005LFinterpretation}.
Early merger phases may have enhanced obscuration \citep{Hopkins2005,Hopkins2006OriginModel},
both galaxy-scale (Compton-thin) and nuclear (Compton-thin and Compton-thick).
Additionally, a substantial part of the faint AGN population is probably
associated with secular triggering mechanisms \citep[e.g.][]{Treister2012}.
Based on these considerations, one may therefore argue for an evolutionary
SMBH ``life'' consisting of {[}illustrative time share in brackets{]}:
(1) no accretion, and therefore no AGN detection {[}89\%{]}, (2) nearby
gas leading to obscuration and triggering of a faint AGN {[}10\%{]},
either at the onset of a merger or through secular accretion events,
(3) major merger triggering a bright AGN, which immediately clears
the vertically extended obscurer but shines for a while before fading
{[}1\%{]}. Such a duty cycle would give rise to the observed obscured
fractions (Figure \ref{fig:L-dep-sketch}) while also respecting the
luminosity function of AGN. To summarise, luminous AGN and faint AGN
may live in different environments with different gas reservoirs feeding
them; therefore unifying the obscuration properties in a luminosity-dependent
torus may not be appropriate.

\section{Conclusions}

\label{sec:Summary}Using only observational relations, we predict
the covering fractions of galaxy-scale gas as relevant for the AGN
population. 

Our findings can be summarised as follows:
\begin{enumerate}
\item Galaxy-scale gas does not provide Compton-thick lines of sight. We
therefore conclude that heavily obscured AGN are associated with nuclear
obscuration, and propose the value $\NH=10^{23.5}{\rm cm}^{-2}$ as
a demarcation line for singling out the nuclear obscurer. 
\item Galaxy-scale gas covers substantial fractions of the SMBH population
at $\NH\approx10^{22-23.5}{\rm cm}^{-2}$, sufficient to fully explain
the obscured fraction at the bright end. 
\item We constrain the nuclear obscurer by subtracting the galaxy-scale
obscuration. A Compton-thick covering of $\sim35\%$ is necessary.
The extent of the Compton-thin obscurer depends on both accretion
luminosity and black hole mass. 
\item We present the empirical \PuffedTorus obscurer model (Section \ref{sub:PuffedTorus}).
It describes the behaviour of the AGN nuclear obscurer (torus) and
galaxy-scale obscurer as a function of accretion luminosity, black
hole mass, host galaxy stellar mass and is consistent with the available
observations. We discuss physical effects which may give rise to this
behaviour. This model can be used as a sub-grid model to post-process
cosmological simulations.
\item We investigated the inside of simulated galaxies from state-of-the-art
hydro-dynamic, cosmological simulations. These verified the assumptions
in our approach by ray-tracing the galaxy gas. Some of these simulations
produce obscured fractions consistent with observations. However,
the results are highly sensitive to the adopted feedback models. We
therefore suggest the Compton-thin obscured AGN fraction as a diagnostic
to rule out feedback models. This diagnostic can be applied already
at early cosmic times ($z=2-3$).
\end{enumerate}
\begin{algorithm}
\caption{\label{alg:code}Python code for generating a column density under
the radiation-lifted torus model. Returns for a specific AGN with
a given X-ray luminosity and black hole mass either a unobscured,
Compton-thin obscured or Compton-thick obscured column. The parameters
are described in Section~\ref{sub:PuffedTorus} and listed in Table~\ref{tab:Parameters}.
The relations of \citet{Marconi2004} are useful to convert bolometric
accretion luminosities to intrinsic $2-10\text{\,keV}$ X-ray luminosity. }

\begin{lstlisting}[language=Python,basicstyle={\footnotesize},tabsize=4]
from numpy import exp, random
p     = 0.666
width = 0.5
fCT   = 0.35
fhost = 0.4
fnuc  = 0.3 / (1 - fhost)

def generate_logNH(logLx, logMBH, logMstar=None):
	if random.uniform(0, 1) < fCT:
		return random.uniform(24, 26)
	logLpeak = (logMBH - 7.87) * p + 43
	fhost = 0.4
	if logMstar is not None:
		NH_host = 21.7 + (logMstar - 9.5)*0.38
		if random.gauss(NH_host, 0.5) > 22:
			return random.uniform(22, 24)
	elif random.uniform(0, 1) < fhost:
		return random.uniform(22, 24)
	delta = (logLx - logLpeak) / width
	fobsc = fnuc * exp(-0.5 * delta**2)
	if random.uniform(0, 1) < fobsc:
		return random.uniform(22, 24)
	else:
		return random.uniform(20, 22)
\end{lstlisting}
\end{algorithm}

\section*{Acknowledgements}

We thank the anonymous referee for useful comments. JB thanks Antonis
Georgakakis and Dave Alexander for insightful conversations. JB thanks
Joop Schaye, Matthieu Schaller and Yetli Rosas-Guevara for detailed
comments on the analysis of hydrodynamic simulations. JB thanks Klaus
Dolag, Sergio Contreras and Torsten Naab for conversations about hydro-dynamic
simulations. 

We acknowledge support from the CONICYT-Chile grants Basal-CATA PFB-06/2007
(JB, FEB), FONDECYT Regular 1141218 (FEB), FONDECYT Postdoctorados
3160439 (JB), ``EMBIGGEN'' Anillo ACT1101 (FEB), and the Ministry
of Economy, Development, and Tourism's Millennium Science Initiative
through grant IC120009, awarded to The Millennium Institute of Astrophysics,
MAS (JB, FEB). 

\bibliographystyle{mnras}
\bibliography{agn,grb,sim}

\appendix

\section{The impact of sub-grid physics on simulated covering fractions}

\label{sec:subgrid-discuss}

We discuss three aspects which affect the results: (1) different sub-grid
physics, most notably stronger feedback mechanisms, (2) differences
between active and passive galaxies, (3) unresolved substructure of
the gas.

The strength of EAGLE is that we can explore how variations of the
physics affect the results. We focus on the fraction of AGN with $\NH>10^{22}\text{cm}^{-2}$,
and compare with EAGLE physics variations in the top panel of Figure~\ref{fig:gasobscprofile-many}.
Recent large-scale AGN surveys \citep[e.g.][]{Ueda2014,Buchner2015}
find a intrinsic fraction of about $\sim40\%$, with uncertainties
encompassing the region $25-65\%$ (dotted vertical lines). We consider
any simulation with fractions outside as ruled out by observations.
At $z=1$ (crosses), this nicely separates the EAGLE physics variations
into two groups, one close to the observational constraints, and one
significantly over-predicting the fraction of obscured AGN. The EAGLE
reference simulation in a large cosmological volume (\texttt{L0100N1504\_REFERENCE})
belongs to the latter group, as do the simulation runs in medium-size
volumes (\texttt{L0025N0752\_REFERENCE}, \texttt{L0025N0752\_RECALIBRATED}).
We now investigate which changes make the simulation agree better.
\citet{Crain2015} presents these variations in detail.
\begin{figure}
\begin{centering}
\includegraphics[width=1\columnwidth]{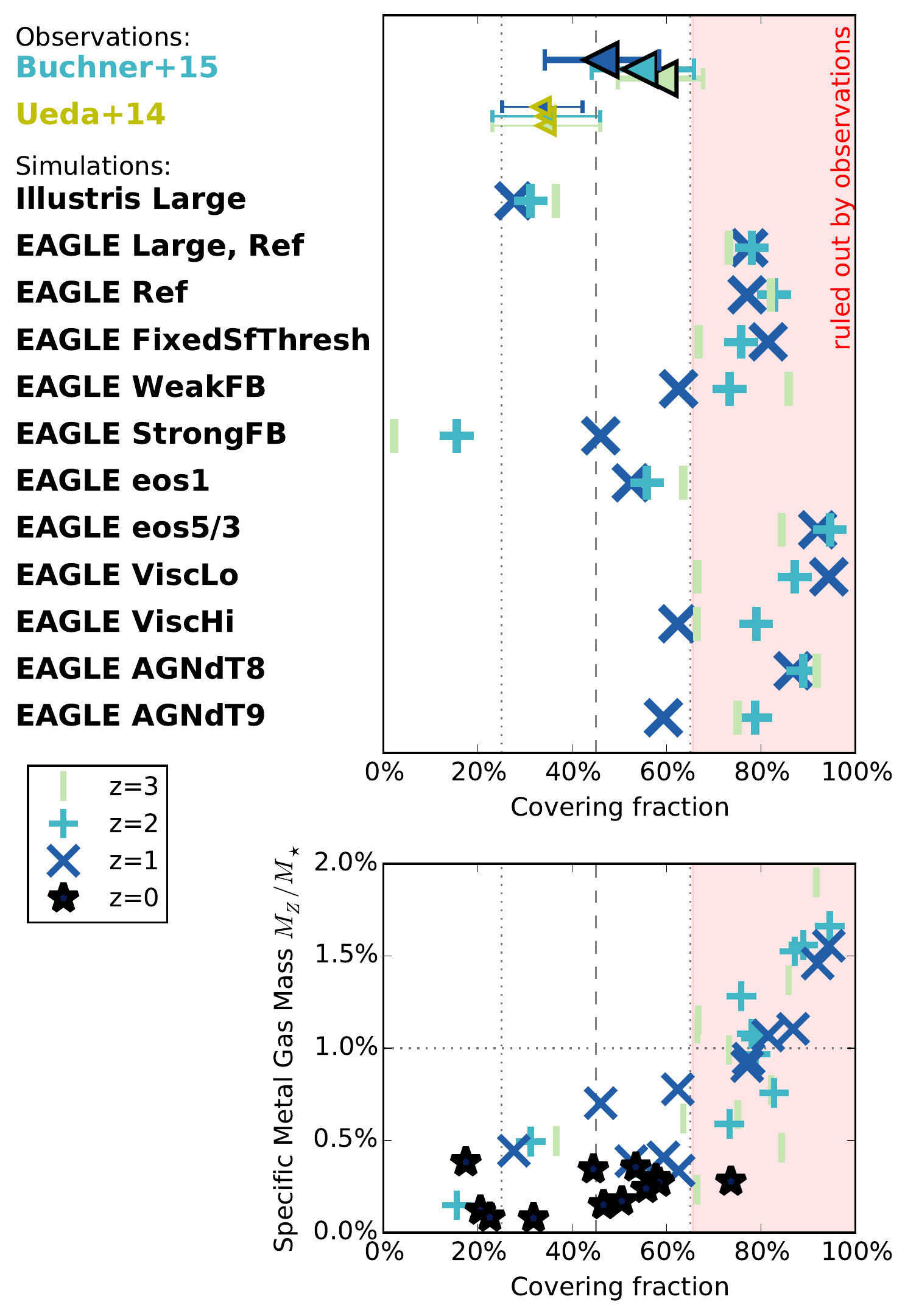}
\par\end{centering}

\caption[Obscured fractions of various simulations]{\label{fig:gasobscprofile-many}Obscured fractions from various simulations.
\emph{Top panel}: We compare the fraction of obscured, Compton-thin
AGN at $z=1-3$ from observations (left-pointing triangles) to various
cosmological simulations. Vertical lines indicate covering fractions
of 25\% (dotted), 45\% (dashed) and 65\% (dotted). Only some simulations
consistently fall within the dotted lines. \emph{Bottom panel}: The
obscured fraction correlates with the average metal gas fraction of
galaxies. Only simulations with specific metal gas fractions of $M_{{\rm Z}}/M_{\star}<1\%$
yield covering fractions consistent with observations.}
\end{figure}

Star formation-related feedback (supernovae, stellar winds, radiation
pressure, cosmic rays) was altered in the WeakFB and StrongFB models.
The efficiency threshold was modified by factors of $0.5$ and $2$
respectively, relative to the reference model. Here we find, surprisingly,
that both models produce lower obscured fractions than the reference
model. Strong feedback leads to underproduction of massive galaxies
at $M_{\star}>10^{10.5}M_{\odot}$ \citep[their Figure 10]{Crain2015},
thereby biasing the galaxy population to small gas masses. It is less
clear why the WeakFB produces a small obscured fraction. Presumably
the feedback is not sufficient to vertically puff up galaxies, thereby
reducing covering fractions. However, these variations are ruled out
by other observations, e.g. the SMF \citep[their Figure 10]{Crain2015}.

Next we discuss the effects of feedback from accreting SMBHs. In the
EAGLE simulation, three parameters affect the triggering, efficiency
and impact of AGN feedback respectively. The equation of state of
the ISM can be modified from its reference value (4/3) to $1$ (\texttt{eos1}).
The increased sound speed then increases the accretion onto black
holes in the simulation, which increases AGN feedback. Once near the
black hole, matter is placed into the black hole with a Bondi accretion
formula modified by a viscosity parameter. Increasing the viscosity
(\texttt{ViscHi}) allows gas to loose angular momentum and accrete
more efficiently. Once accreted, the temperature of particles near
AGN is increased by $\Delta T=10^{8.5}{\rm K}$, stochastically, bringing
the gas into the metal cooling regime. \citet{Schaye2015} test the
impact of increasing this temperature to $\Delta T=10^{9}{\rm K}$
(\texttt{AGNdT9}). Each of these three modifications (\texttt{eos1},
\texttt{ViscHi}, \texttt{AGNdT9}) leads to a reduction of the obscured
fraction of AGN (see Figure~\ref{fig:gasobscprofile-many}), while
contrary modifications (\texttt{eos5/3}, \texttt{ViscLo}, \texttt{AGNdT8})
do the opposite. It is worth noting that \citet{Schaye2015} preferred
the \texttt{AGNdT9} variation over the reference simulation because
it better fits soft X-ray observations of cluster gas. We also point
out that Illustris uses relatively strong feedback, as it implements
three different AGN feedback schemes (thermal, kinetic and radiation),
whereas EAGLE uses only stochastic heating.

AGN outflows or radiation pressure may decrease the covering fractions
momentarily. We have so far considered \emph{all} simulated galaxies
and found that they produce high covering fractions. Arguably these
fractions are consistent with the covering fraction of low-luminosity
AGN. Therefore, one could propose that AGN feedback at high luminosities
modifies the galaxy in such way that covering fractions are reduced.
Apriori this proposal appears unlikely, because nuclear gas should
be affected first. Additionally, studies comparing the morphology
of active and passive galaxies have found little evidence that these
are different, by comparing appearances with asymmetry and concentration
measures \citep{Grogin2003,Grogin2005,Pierce2007,Gabor2009,Kocevski2012}
or visual classification \citep{Kocevski2012}. Indeed, our results
remain unchanged if in the EAGLE reference simulation we only consider
simulated galaxies with instantaneous black hole accretion rates corresponding
to $L(2-10\,{\rm keV})>10^{42}{\rm erg/s}$, assuming a radiative
efficiency of 10\% and bolometric corrections of \citet{Marconi2004}.
In fact, because active galaxies are preferentially gas-rich, star-forming
galaxies in that simulation, the average column density is higher
by a factor of 2, which increases the discrepancy.

A clumpy ISM may decrease the covering fractions. For instance the
galactic ISM is structured into parsec-size clumps with filling factors
of $1\%$ \citep{Cox2005}. Such clumps could not be resolved by simulations.
However, as a LOS passes through large distances of the ISM ($1-\text{few}$
kiloparsecs), this clumpiness averages out. Additionally, clumpiness
would effectively only redistribute the obscured fraction to both
lower and higher column densities, potentially violating the constraints
of higher column densities. There are also differences in the hydrodynamics
code schemes and their accuracy. However, these are less important
than the chosen sub-grid models \citep[J. Shaye, priv. comm., see][]{Scannapieco2012,Schaller2015,Cui2016,Sembolini2016}
in the present non-classical SPH simulations.

To summarise, our obscured fraction diagnostic is a highly sensitive
test of feedback recipes. It can easily rule out feedback models already
at early times (e.g. $z=2-3$) in the simulations, if they produce
very high fractions of obscured AGN. This effectively places limits
on the metal gas mass content of galaxies at $z=0-3$, as the bottom
panel of Figure~\ref{fig:gasobscprofile-many} demonstrates. There
we plot the average specific metal gas mass $M_{{\rm Z}}/M_{\star}$
that is bound to the galaxy and within $10\,{\rm kpc}$. The average
was computed by fitting a powerlaw at $M_{\star}>10^{9.5}M_{\odot}$
with $3\sigma$ clipping, and evaluating $M_{{\rm Z}}/M_{\star}$
at $M_{\star}=10^{10}M_{\odot}$. Adopting a moving average median
instead gives comparable results. We find that the specific metal
gas mass correlates well with the obscured fraction. The fact that
unobscured AGN are found, i.e. $f_{\text{obsc}}\leq65\%$, implies
that for the present set of simulations we can place a limit on the
specific metal gas mass of $M_{{\rm Z}}<1\%\cdot M_{\star}$, for
the redshift range $z=0-3$.

\section{The gas mass inside galaxies}

\begin{figure*}
\begin{centering}
\includegraphics[width=1\textwidth]{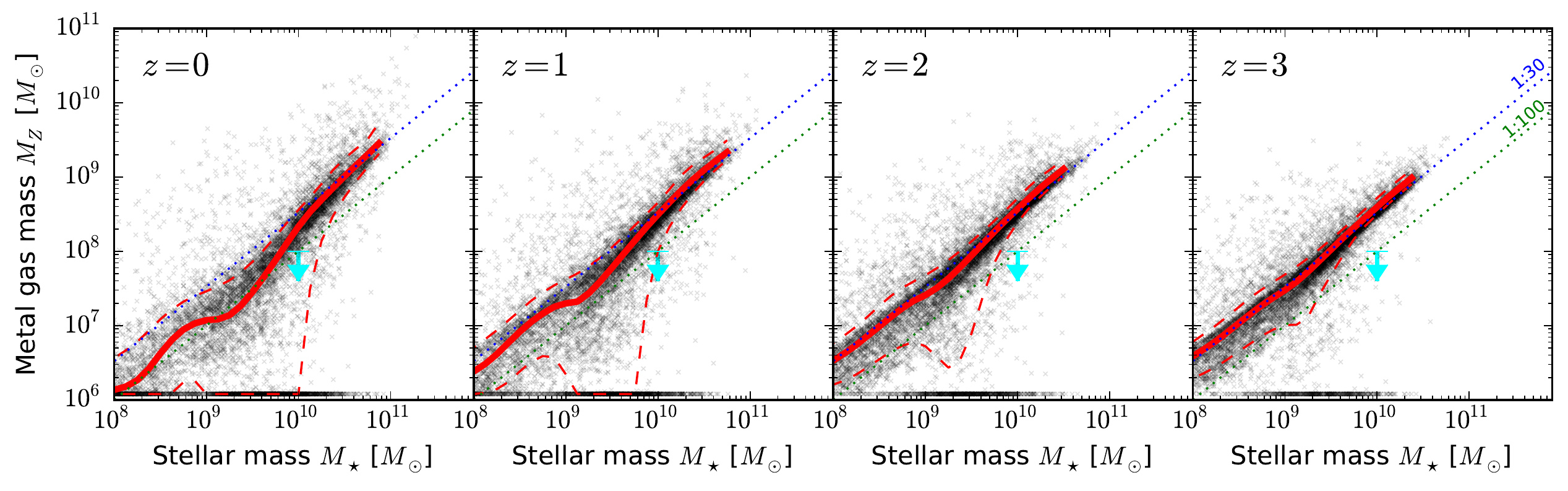}
\par\end{centering}

\caption{\label{fig:MZ-sam}The metal gas mass in galaxies according to the
SAGE semi-analytic model. The solid red curve indicates the median,
while dashed red curves are the $1\sigma$-equivalent quantiles. The
ratio of metal gas to stellar mass lies between $1:30$ and $1:100$
(dotted lines). The magenta upper limit is derived in Appendix~\ref{sec:subgrid-discuss}
for the mass of the central $10\,{\rm kpc}$.}
\end{figure*}
\label{sub:semi-analytic}Important constraints on how much gas resides
in galaxies can be drawn from cosmological simulations. Such simulations
evolve the matter density available at the Big Bang into collapsing
bound structures. Semi-analytic models successfully reproduce many
features of galaxies, including the stellar mass function of galaxies
and their colour distribution \citep{Croton2006,Somerville2008,Hirschmann2012,Fanidakis2012}.
As an illustrative case, we consider the model of \citet{Croton2016}.
Figure \ref{fig:MZ-sam} plots the metal gas mass (the X-ray obscurer)
present as a function of galaxy stellar mass. The median (red curve)
falls consistently in the 1:30 to 1:100 range for the ratio of metal
gas mass to stellar mass. With AGN host galaxies primarily drawn around
the $M_{\star}=10^{10-11}M_{\odot}$ regime, the total gas available
to obscure a central point source is about $M_{\text{Z}}\sim10^{9}M_{\odot}$.
The Illustris simulation shows very similar results at $z=1-3$ obeying
the same gas ratios. However at $z=0$, the gas content at the high-mass
end is reduced because of the strong feedback implemented in that
simulation.

We present a simple calculation to show that Compton-thick column
densities (i.e. $\NH>1.5\times10^{24}\text{cm}^{-2}$) cannot be achieved
by accumulating the galaxy gas over several kpc. In X-ray spectral
analysis, the equivalent hydrogen column density $\NH$ is usually
computed assuming solar abundances. To mimic this, we convert the
metal mass in particles to the number of hydrogen atoms assuming solar
mass fractions of the nearby ISM from \citet{Wilms2000}:

\begin{eqnarray*}
n_{\text{H}} & = & \left.\frac{f_{\text{X}}}{f_{\text{Z}}}\right|_{\text{solar}}\times\frac{\rho_{\text{Z}}}{m_{\text{H}}}
\end{eqnarray*}
Inserting numbers including the hydrogen mass $m_{\text{H}}$ we find
\begin{eqnarray*}
n_{\text{H}} & = & 0.737\times10^{22}\text{cm}^{-2}\frac{1}{1\text{kpc}}\times\frac{\rho_{\text{Z}}}{10^{6}M_{\odot}\text{kpc}^{-3}}.
\end{eqnarray*}
For example, a 1kpc ray in a region of metal gas density $10^{6}M_{\odot}/\text{kpc}^{3}$
results in a measured column density of $\NH\approx10^{22}\text{cm}^{-2}$. 

The gas inside a galaxy may be arranged in a multitude of ways to
achieve a covering with column density $\NH$. If we consider only
gas \emph{outside} a certain radius $R$, the most effective obscurer,
i.e. the one with the least mass but complete covering, is an infinitely
thin shell at that radius $R$. Its mass is easily computed as 
\[
M_{\text{H}}(\NH,>R)=4\pi\cdot R^{2}\times\NH\times m_{\text{H}}.
\]
Converting to metals using the factor $\left.\frac{f_{\text{Z}}}{f_{\text{X}}}\right|_{\text{solar}}$and
expressing in conventional units, this limit is
\[
M_{\text{Z}}(\NH,>R)=2.6\times10^{9}M_{\odot}\cdot\frac{\NH}{1.5\times10^{24}\text{cm}^{-2}}\cdot\left(\frac{R}{1\text{kpc}}\right)^{2}.
\]
\begin{figure}
\includegraphics[width=1\columnwidth]{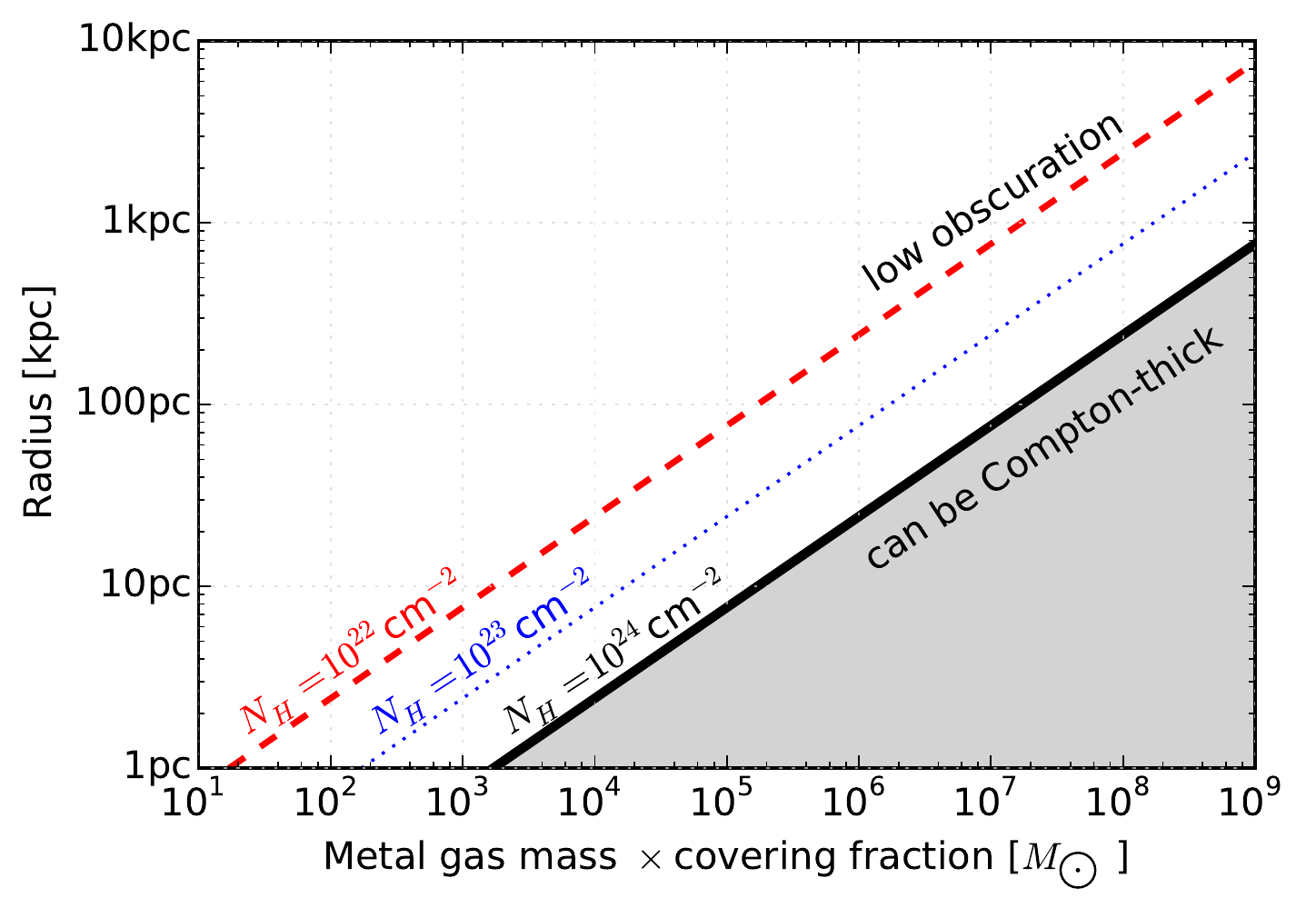}

\caption{\label{fig:sim-MZ-obsc}X-ray obscuration that a given metal gas mass
can reach. An obscurer of column density $\NH$ outside a radius $R$
has a minimum metal gas mass limit $M_{Z}$ (lines). For example,
a $M_{Z}=10^{9}M_{\odot}$ gas mass has to be brought withing 100pc
to completely enshroud a region with Compton-thick column densities.
Keep in mind that such masses are only found in very massive galaxies
($M_{\star}>3\cdot10^{10}M_{\odot}$, see Figure \ref{fig:MZ-sam}).}
\end{figure}
Therefore, a metal mass larger than $2.6\times10^{9}M_{\odot}$ is
required to create a Compton-thick obscurer outside the central $1\text{kpc}$.
Or equivalently, a metal mass of $2.6\times10^{9}M_{\odot}$ has to
be brought to the central $\text{kpc}$ to act as a Compton-thick
obscurer. Note that this mass limit scales with the covering factor;
for example obscuration of $10\%$ of the sky requires $10\%$ of
the mass. This simple limit is shown for several levels of obscuration
in Figure \ref{fig:sim-MZ-obsc}. \citet{Risaliti1999} already noted
that such large masses at radii further than a few $10{\rm pc}$ are
ruled out in NGC1068 and Circinus, because they would gravitationally
dominate the central region.

Given that metal gas masses are below $2.6\times10^{9}M_{\odot}$
in all but the most extreme galaxies (Figure~\ref{fig:MZ-sam}),
it follows that galaxy gas, when outside the central $1\text{kpc}$,
is incapable of Compton-thick obscuration with substantial covering
angles. In other words, some of that mass has to reside inside the
central $1\text{kpc}$ to create Compton-thick covering. Admittedly,
this is a weak constraint. However the result holds independently
of the geometry of the gas, the galaxy type and is also applicable
to mergers. As an example, we consider a $M_{\star}=10^{9}M_{\odot}$
galaxy merging into a $M_{\star}=10^{10}M_{\odot}$ galaxy (minor
merger). Furthermore, lets assume that all of its gas ($M_{Z}\approx10^{7}M_{\odot}$)
is made available as an obscurer. From Figure~\ref{fig:sim-MZ-obsc},
we infer that the entire amount of gas must land within $100{\rm pc}$
of the AGN in order to completely enshroud the SMBH in Compton-thick
columns. More quantitative conclusions depend on the geometrical distribution
of the gas in the galaxy. We analyse the galaxies produced by hydrodynamic
simulations in Section \ref{sec:Cosmological-simulations}.
\end{document}